\documentclass[pra,a4paper,twocolumn,superscriptaddress,longbibliography,nofootinbib]{revtex4-2}

\usepackage{amssymb}
\usepackage{amsmath}
\usepackage{amsfonts}
\usepackage{graphicx}
\usepackage{bm}
\usepackage[dvipsnames]{xcolor}
\usepackage{multirow}
\usepackage{comment}
\usepackage{subfigure} 
\usepackage[colorlinks]{hyperref}
\usepackage{transparent}

\usepackage{xcolor}
\usepackage{cancel}

\usepackage{printlen}
\uselengthunit{pt}

\DeclareMathOperator{\sgn}{sgn}
\DeclareMathOperator{\re}{Re}
\DeclareMathOperator{\im}{Im}

\renewcommand{\a}{\alpha}
\newcommand{\g}{\gamma}

\newcommand{\ua}{\uparrow}
\newcommand{\da}{\downarrow}
\newcommand{\D}{\Delta}
\newcommand{\pd}{\partial}
\renewcommand{\l}{\lambda}

\begin{document}

\title{Majorana bound states in chiral ferromagnet--superconductor heterostructures revisited}

\author{A. S. Slobodskoi}

\affiliation{Moscow Institute for Physics and Technology, 141700, Moscow, Russia}

\affiliation{\hbox{L. D. Landau Institute for Theoretical Physics, Semenova 1-a, 142432, Chernogolovka, Russia}}

\author{S. S. Apostoloff}

\affiliation{\hbox{L. D. Landau Institute for Theoretical Physics, Semenova 1-a, 142432, Chernogolovka, Russia}}

\affiliation{Laboratory for Condensed Matter Physics, HSE University, 101000 Moscow, Russia}

\author{I. S. Burmistrov}

\affiliation{\hbox{L. D. Landau Institute for Theoretical Physics, Semenova 1-a, 142432, Chernogolovka, Russia}}

\begin{abstract}
Majorana zero modes are central to the pursuit of fault-tolerant topological quantum computation. While traditionally sought in one-dimensional hybrid nanowires, a robust alternative platform involves heterostructures combining superconductors with noncollinear magnets. This work focuses on a particularly promising system: a chiral ferromagnet hosting a magnetic skyrmion coupled to a superconducting film containing a superconducting vortex. Such skyrmion-vortex pairs have recently been realized experimentally and are theorized to harbor localized Majorana states, offering a potential pathway for braiding operations. We present a comprehensive theoretical analysis of the low-energy quasiparticle bound states in these heterostructures. Extending previous studies, we develop an analytical framework for the Majorana wavefunctions as well as the wavefunctions and spectrum of other lowlying states within a Bogoliubov-de Gennes approach. Our analytical results explicitly demonstrate the critical role of spin-orbit coupling for the stabilization of Majorana modes and provides approximate analytical expressions for low-lying states localized at the vortex, both with and without an accompanying skyrmion. The derived analytical results show excellent agreement with numerical simulations. We further elucidate the role of realistic effects, including vector potentials and texture perturbations from stray magnetic fields, to assess their impact. 
\end{abstract}

%\date{\today (v.5)}

\maketitle

\section{Introduction}

The intense interest in Majorana fermions within condensed matter physics is largely fueled by their potential for fault-tolerant topological quantum computation, as they are predicted to exhibit non-Abelian statistics suitable for quantum information processing~\cite{Kitaev2003,Nayak2008}. The canonical approach to realizing these exotic quasiparticles involves engineering one-dimensional topological superconductors. Original theoretical proposals showed that combining semiconductors with strong spin-orbit coupling, magnetic fields, and conventional $s$-wave superconductors via the proximity effect could effectively induce a topological $p$-wave superconducting phase, hosting localized Majorana zero modes at wire ends or at defects~\cite{Kitaev2001, Oreg2010, Lutchyn2010, Sau2010}. Subsequent experimental efforts in hybrid semiconductor-superconductor nanowires reported signatures of these edge states~\cite{Mourik2012, Das2012, Sun2015, Sato2017,Aghaee2023}, but without conclusive evidence for their existence \cite{DasSarma2023,Frolov2023}. In parallel, a promising alternative direction has emerged, exploring two-dimensional heterostructures. Here, a thin superconducting film is coupled to a magnetic system with a noncollinear spin arrangement or with a collinear spin order together with spin-orbit coupling~\cite{Nakosai2013,Black-Schaffer2013,Chen2015,Mascot2019,Rex2020}. This approach offers several distinct advantages (see Refs. \cite{Zlotnikov2021,Valkov2022} for a review). The proximity-induced topological superconductivity in such systems can be remarkably robust, with Majorana edge states or vortex-bound zero modes emerging over a much wider range of external parameters. 

An example of a magnetic system that exhibits a localized noncollinear magnetic texture and intrinsic spin-orbit coupling is a chiral ferromagnetic film which hosts skyrmions \cite{Bogdanov1989}. These are topologically stable collective excitations in noncentrosymmetric magnets (for a review, see Refs. \cite{BorisovUFN2020en,Tretiakov2021,Petrovich2024}). It has been demonstrated that skyrmions in a chiral ferromagnet -- superconductor heterostructure not only support localized Yu-Shiba-Rusinov states \cite{Pershoguba2016}, but also in some cases, they can host localized Majorana states \cite{Yang2016,Gungordu2018,Garnier2019}. Typically, the existence of Majorana states requires a skyrmion 
with an even topological charge, which is not easy to control experimentally \cite{Hassan2024}. 
However, Ref. \cite{Rex2019} showed that the need for a skyrmion with even topological charge to realize the Majorana state can be elegantly avoided by considering a skyrmion with the unit topological charge coupled to a superconducting vortex. Later, it was shown \cite{Pathak2022} that Majorana states may also exist in coupled magnetic and superconducting vortices.

Several pathways can lead to the creation of a bound skyrmion-vortex pair. When the interface is of high quality, the interplay of spin-orbit coupling and proximity effects can nucleate a skyrmion-vortex pair~\cite{Hals2016,Baumard2019}. Beyond this interface-mediated route, a universal interaction mechanism operates regardless of interface quality, driven by the effect of the vortex's stray magnetic field on the magnetization of the skyrmion~\cite{Dahir2018,Dahir2019,Menezes2019,Dahir2020,Andriyakhina2021,Andriyakhina2022}. The precise outcome of this magnetostatic interaction is governed by the skyrmion's internal spin texture. Notably, for a N\'eel-type skyrmion, the energetically favored state may involve both a stable coaxial configuration as well as locking the vortex and skyrmion centers at a finite distance apart (for a review, see Ref.~\cite{UFN-SK}). 

%%%%%%%%%%%%%%%%%%%%%%%%%%%%
\begin{figure}[t]
\includegraphics[width=\columnwidth]{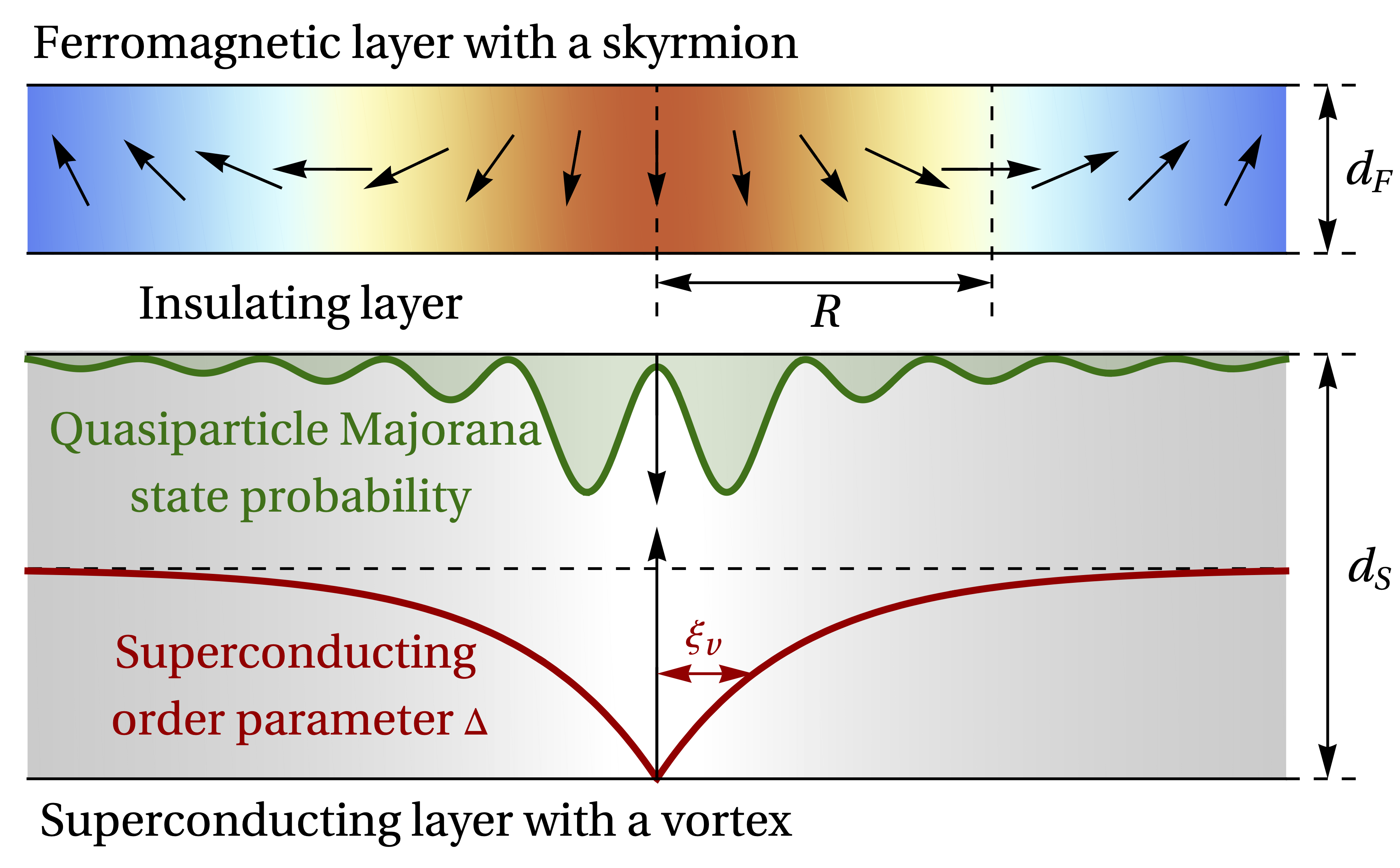}
\caption{A schematic illustration of a heterostructure composed of a chiral ferromagnetic film (FM) and a superconducting film (S) with a skyrmion and a superconducting vortex (see text).}
\label{Fig:setup}
\end{figure}
%%%%%%%%%%%%%%%%%%%%%%%%

Recently, skyrmion-vortex pairs have been observed in chiral ferromagnet--superconductor heterostructures 
\cite{Petrovic2021, Machain2021,Xie2023}. This makes such a pathway for the realization of Majorana states potentially promising. In particular, recent works have proposed and analyzed protocols for 
braiding Majorana states which are hosted by skyrmion-vortex pairs~\cite{Nothhelfer2022,Konakanchi2023}.

An application of Majorana states in skyrmion-vortex pairs for quantum computation requires a detailed understanding of the spatial structure of these states, as well as knowledge of other possible low-lying localized states. Previous works \cite{Rex2019,Pathak2022} have provided only numerical analysis of this problem, but a more detailed analysis is needed. In this paper, we revisit the issue of low-energy bound states in skyrmion-vortex pairs in a chiral ferromagnet--superconductor heterostructure.  

Within the Bogoliubov-de Gennes (BdG) Hamiltonian, we study the low-lying bound quasiparticle states at the interface between a chiral ferromagnetic film and a superconducting film containing a vortex (see Figure~\ref{Fig:setup}). Our work extends the previous study in Ref.~\cite{Rex2019}, and we make several improvements. Under reasonable assumptions, we develop an analytical theory for Majorana wavefunctions as well as the wavefunctions and spectrum of low-lying bound states. First, we derive approximate analytical expressions for Majorana states located at a superconducting vortex, both with and without a skyrmion in the chiral ferromagnet. Our results demonstrate the crucial importance of spin-orbit coupling for the existence of these localized Majorana states. Our analytical findings are in excellent agreement with our numerical simulations. Second, we derive expressions for the wavefunctions and energies of low-lying bound states located at the vortex and the edge of the system, as a function of integer-valued angular momentum. Third, we estimate the role of such realistic effects as the vector potential induced by the vortex and the noncollinear magnetic texture and the perturbation of the magnetic texture due to the stray magnetic field generated by the superconducting vortex.

The outline of the paper is as follows. \color{black} We start with the formulation of the model and derivation of the general condition for the existence of the Majorana states in Sec. \ref{Sec:Model}. The analytic theory and numerical results for the wavefunction of the Majorana state localized at the superconducting vortex without and with a coaxial skyrmion are presented in Sec. \ref{Sec:Core}. In Sec. \ref{Sec:Edge} we discuss the structure of the Majorana state localized near the edge of the system. The analytic and numerical results for the wavefunction and energies of states with nonzero angular momentum localized at the superconducting vortex are given in Sec. \ref{Sec:finL}. We end the paper with discussions and conclusions in Sec. \ref{Sec:Disc}. Some technical details are delegated to the Appendix.

\section{The model\label{Sec:Model}}

\subsection{Formalism}

We focus on the following quasiparticle Hamiltonian at the interface between a chiral ferromagnet and a superconducting film \cite{Yang2016,Rex2019,Pathak2022}:
\begin{gather}
H=\frac{\tau_z}{2m}(-i\nabla-e \bm{A}\tau_z)^2-\mu \tau_z + \tau_x \re \Delta  - \tau_y \im \Delta   \notag \\
 {}+J \bm{\sigma} \bm{m}   + \alpha \tau_z [\bm{\sigma}\times (-i \nabla-e \bm{A}\tau_z)]_z .
\label{eq:H:1}
\end{gather}
Here, $\sigma_j$ and $\tau_j$ represent the Pauli matrices operating in the spin and Nambu spaces, respectively. The chemical potential is denoted by $\mu$ and the quasiparticle mass is denoted by $m$. 
The antiferromagnetic exchange coupling~$J>0$ 
controls the Zeeman effect on the quasiparticles caused by the magnetization~$\bm{m}$ of the chiral ferromagnet. The spin-orbit coupling is in the standard Rashba form, with a magnitude denoted by $\alpha$. 

The superconducting order parameter, denoted by $\Delta(\bm{r})=\Delta(r)e^{i\varphi}$, has a non-trivial phase. Here $\varphi$ is the polar angle and $r=|\bm{r}|$ is the radial coordinate. This choice of the order parameter represents the presence of a superconducting vortex at the origin, $\bm{r}=0$, with magnetic flux pointing in the negative $z$ direction (here the $z$-axis is perpendicular to the interface), see Fig.~\ref{Fig:setup}. In general, the order parameter profile $\Delta(r)$ must be determined self-consistently, see e.g. Ref. \cite{Bardeen1969,Gygi1991}. For a sake of simplicity, we will  consider a standard approximation for the order parameter near the superconducting vortex, given by: 
$\Delta(r) = \Delta_{\infty}[1-\exp(-r/\xi_v)]$, where 
$\Delta_\infty>0$ 
is the magnitude of the order parameter away from the vortex. A magnitude of the length scale $\xi_v$ typically  lies between the Fermi wavelength and the superconducting coherence length.  

The Hamiltonian~\eqref{eq:H:1} includes the electro-magnetic vector potential, $\bm{A}= \bm{A}^{v}+\bm{A}^{m}$, which is composed of two parts: the vector potential produced by a vortex, $\bm{A}^{v}$, and the vector potential created by the magnetic texture, $\bm{A}^{m}$. The vector potential above the superconducting film (at $z = 0$) created by a single vortex located at the origin is expressed as~\cite{Carneiro2000}:
\begin{eqnarray}
	{\bm A}^{v}(r)  =  -\bm{e}_{\phi} \phi_0  
	\int_0^{\infty} \frac{dq}{2\pi} \frac{\mathcal{J}_1(qr)}{F(q)},
	\label{eq:vortex:A}
\end{eqnarray}
where $\phi_0 = h/2|e|$ is the magnetic flux quantum and $\mathcal{J}_k(x)$ stands for the Bessel function of the first kind.
Here and below, $\bm{e}_{r}$, $\bm{e}_{\phi}$, and $\bm{e}_{z}$ are unit vectors in the radial, azimuthal, and axial directions of the cylindrical coordinate system.  
For an arbitrary width of the superconducting film, $d_S$, the function $F(q)$ is given by
\begin{equation}
	F(q) = \lambda_L^2 \kappa[ q+ \kappa-( \kappa-q)e^{- \kappa d_S}][1-e^{-\kappa d_S}]^{-1},
	\label{eq:SM_Fq}
\end{equation}
where $\lambda_L$ is the London penetration length and ${\kappa = \sqrt{q^2 + \lambda_L^{-2}}}$. 
In what follows, we assume that the London penetration length is larger than the vortex size, $\lambda_L\gg \xi_v$.
In the limit of a thin film, $d_S \ll \lambda_L$, and when we are not too close to the center of the vortex, $r \gg d_S$, we can approximate the integral in Eq.~\eqref{eq:vortex:A} using $F(q)  \simeq  1 + 2q\lambda$, where $\lambda=\lambda_L^2/d_S$ is known as the Pearl length \cite{Pearl1964}. 
Note that formally Eq.~\eqref{eq:vortex:A} works only for $r\gg\xi_v$ and should be corrected for $r\lesssim\xi_v$. However, as we see further that is not significantly affect the Majorana states.

The vector potential $\bm{A}^{m}$, induced by the magnetization of a ferromagnetic film with width $d_F$ can be determined using the Maxwell-London equation:
\begin{gather}
\nabla\times(\nabla\times\bm{A}^{m})+\lambda_L^{-2}\Theta(-z)\Theta(z+d_S)\bm{A}^{m} \notag \\
=4\pi M_s
\Theta(z)\Theta(d_F-z)\nabla\times \bm{m}  .
\end{gather}
Here $M_s$ is the saturation magnetization. In this work we will focus on the magnetic texture in the form of the N\'eel-type skyrmion, 
\begin{equation}
    \bm{m}(\bm{r}) = \cos\theta(r) \bm{e}_z + \sin  \theta(r) \bm{e_r} .
\label{eq:MM}
\end{equation}
We note that the above expression is written in the polar coordinate system with the origin at the center of the skyrmion. The angle $\theta(r)$ can be found from the solution of the magnetic problem (see Ref. \cite{UFN-SK} for a review). In this work we consider the angle $\theta(r)$ to be a given function. 

For the most part of the paper we assume that magnetization~$\bm{m}$ is directed in the $z$-direction far from the vortex, $\bm{m}\to\bm{e}_z$ at $r\to\infty$. In particular, $\theta(r)\equiv 0$ corresponds to a film  homogeneously magnetized in $z$-direction. We discuss shortly the opposite case with homogeneous $\bm{m}=-\bm{e}_z$, or $\theta(r)\equiv \pm \pi$, in subsection~\ref{sec:theta_pi}, which is profitable for the case of the skyrmion.

We note that the presence of a superconducting vortex inevitably induces an inhomogeneous magnetization near the origin in the ferromagnet~\cite{Apostoloff2024}. Unlike a skyrmion, this magnetic texture is topologically trivial. 
The vector potential induced by texture~\eqref{eq:MM} at the interface, $z=0$, is as follows \cite{Andriyakhina2021}
\begin{gather}
	{\bm A}^{m}(r)  = \bm{e}_{\phi} 4\pi M_s
	\int_0^{\infty} dq \frac{\mathcal{J}_1(qr) G(q) (1-e^{-q d_F})}{\tilde{F}(q)} , \notag \\
  \tilde{F}(q)=\frac{(\kappa+q)^2-(\kappa-q)^2e^{-2\kappa d_S}}{(\kappa+q)+(\kappa-q)e^{-2\kappa d_S}}
	\label{eq:vortex:M}
\end{gather}
In the thin film limit, $d_S\ll\lambda_L$ the function $\tilde{F}$ simplifies as $\tilde{F}(q)\simeq 2q +1/\lambda$.
The function $G(q)$ is specified by the skyrmion angle and is given as
\begin{equation}
  G(q) = \int_0^\infty dr\, r \mathcal{J}_1(qr) [q+\partial_r\theta(r)]\sin\theta(r)  .   
\end{equation}

We note that the center of the skyrmion can be located at a non-zero distance $a$ from the center of a superconducting vortex. In this case, the Hamiltonian~\eqref{eq:H:1} lacks the polar symmetry. 
In this paper we consider the case of $a=0$. More complicated situation of a skyrmion shifted  with respect to a vortex will be published elsewhere.

\subsection{Existence of the Majorana states}

The Hamiltonian~\eqref{eq:H:1} acts on a four component wave function $\Psi=\{u_\uparrow,u_\downarrow,v_\downarrow,-v_\uparrow\}^T$ and we are interested in the solution of the following stationary BdG equation
\begin{equation}
H \Psi = E \Psi .
\label{eq:2}
\end{equation}
We note that the Hamiltonian~\eqref{eq:H:1} possesses the particle-hole symmetry
\begin{equation}
C H C = - H, \qquad C= \sigma_y\tau_y K ,
\label{eq:C:sym}
\end{equation}
where $K$ is the complex conjugation. The particle-hole symmetry~\eqref{eq:C:sym} implies that if $\Psi_E$ is an eigen state of $H$ with energy $E$ then the state $C\Psi_E$ will be an eigen state with the energy $-E$. 
This fact suggests  that if Eq.~\eqref{eq:2} has a solution for $E=0$, then it might be  
doubly degenerate, with the eigen functions 
$\Psi_0$ and $C\Psi_0$, where $H \Psi_0=0$. 
It is important to note that these states are not the same. Therefore, we can construct their linear combinations:
\begin{equation}
    \Psi^{\pm} = \frac{\Psi_0\pm C\Psi_0}{\sqrt{2}}, \quad C\Psi^\eta = \eta \Psi^\eta, \quad \eta =\pm.
    \label{eq:3}
\end{equation}
These states are what we will call \textit{Majorana states}~\cite{Yang2016}. For these states, we note 
the following relation between the electron-like and hole-like components of the bi-spinor: $u_{\uparrow,\downarrow}=\eta v^*_{\uparrow,\downarrow}$.

In the case where the center of a skyrmion is located at the center of a vortex, we are dealing with the centrally symmetric problem. It is worth introducing the operator for the 
$z$-projection of the generalized angular momentum \cite{Rex2019}:
\begin{equation}
L_z= -i\partial_\varphi +\frac{1}{2}\sigma_z-\frac{1}{2}\tau_z 
\end{equation}
with integer eigen values $l$. The operator $L_z$ commutes with the Hamiltonian,  $[L_z,H]=0$. 
We emphasize that the operator $L_z$ retains its meaning even in the absence of a skyrmion, that is, for $\theta\equiv 0$ or $\theta\equiv \pi$.
The operator of the generalized angular momentum allows us to use the following representation for the eigenfunction:
\begin{equation}
\Psi(r,\varphi) = \sum_{l\in \mathbb{Z}} e^{i \varphi(l -\sigma_z/2+\tau_z/2)} \Phi_l(r) 
\end{equation}
and to reduce the eigenvalue problem~\eqref{eq:2} to the following one
\begin{equation}
    H^{(l)} \Phi_l = E_l \Phi_l .
    \label{eq:5.0}
\end{equation}
Here the Hamiltonian for the $l$-th angular harmonics reads
\begin{align}
H^{(l)} & = 
-\frac{\tau_z}{2m}\left [\Bigl (\partial_r^2 +\frac{1}{r}\partial_r\Bigr ) 
- \Bigl (\frac{\sigma_z{-}\tau_z{-}2l}{2r}+e A_{\varphi}(r)\tau_z\Bigr )^2\right ] 
\notag \\
& -\mu \tau_z + \Delta(r) \tau_x  + J \sin\theta(r)\sigma_x +J \cos\theta(r)\sigma_z 
\notag \\
& + i \alpha \tau_z\sigma_y \left (\partial_r+\frac{1}{2r}\right ) + \alpha \sigma_x\Big[ \frac{1+2l\tau_z}{2r} 
- e A_\varphi(r) \Big].
\label{eq:Hl} 
\end{align}
We note the existence of the following symmetry relations 
\begin{equation}
    C H^{(-l)}C=-H^{(l)}, \qquad C L_z C = - L_z .
\end{equation}
They imply that if we have an eigenstate $\Phi_l$ of the Hamiltonian $H^{(l)}$ with energy $E_l$, corresponding to some given $l$, then the state $C \Phi_l$ will be an eigenstate of the operator $L_z$ with  eigenvalue $-l$ and also of the Hamiltonian $H^{(-l)}$ with energy $E_{-l}=-E_l$. Because the Majorana states are the eigenstates of $H$ with zero energy and belong to the subspace with a given angular momentum $l$, and those eigenfunctions $\Psi^\eta$ and $C\Psi^\eta$ are proportional to each other, cf. Eq.~\eqref{eq:3}, they  
correspond to $l=0$.
Therefore, for the existence of Majorana states, it is  crucial that the $z$-projection of the generalized angular momentum must have integer eigenvalues \cite{Rex2019}. This is achieved through the presence of the spin Pauli matrices in the Hamiltonian~\eqref{eq:H:1}. 
We note that, in general, there are possible zero-energy states with some $l$ and $-l$ ($l>0$). However, these states are not commonly referred to as Majorana states.

Using Eq.~\eqref{eq:3}, we present the solution of Eq.~\eqref{eq:5.0} for $l=0$ and $E_l=0$ as 
$\Phi_0^\eta = \{\phi^\eta,i\eta \sigma_y \phi^\eta\}^T$, where ${\phi^\eta  = \{u_\uparrow,u_\downarrow\}^T}$. Here, we assume that $u_{\uparrow,\downarrow}$ are real functions normalized such that 
\begin{equation}
\int\limits d^2\bm{r} \, (u_{\downarrow}^2+u_{\uparrow}^2)=\dfrac{1}{2} .
\label{eq:norm:cond:m}
\end{equation}

Therefore, we
reduce the eigenvalue problem for the Majorana state to the following one
\begin{equation}
    H^{\eta} \phi^\eta = 0,
    \label{eq:eq:M}
\end{equation}
where $2\times 2$ non-Hermitian Hamiltonian reads
\begin{align}
H^\pm = &-\frac{1}{2m}\Bigl (\partial_r^2 +\frac{1}{r}\partial_r\Bigr ) 
+\frac{1}{2m} \Bigl [\frac{\sigma_z-1}{2r}+e A_{\varphi}(r)\Bigr ]^2 
\notag \\
& -\mu \pm i \Delta(r) \sigma_y  + J \sin\theta(r)\sigma_x +J \cos\theta(r)\sigma_z
\notag \\
& +i \alpha \sigma_y \left (\partial_r+\frac{1}{2r}\right )  + \alpha\sigma_x \Big[\frac{1}{2r} - e A_\varphi(r)\Big] .
\label{eq:3:HH}
\end{align}

\subsubsection{The Majorana state localized at the vortex}

Now we discuss the region of existence of Majorana solutions of Eq.~\eqref{eq:eq:M}. 
We assume that the vector potential $A_\varphi(r)$ and the magnetic texture angle $\theta(r)$ decay fast enough as $r\to \infty$. 
We start from the solution localized at the origin (the center of the superconducting vortex). Detailed analysis of Eq.~\eqref{eq:3:HH} (see Appendix~\ref{App:A}) at $r\to 0$ demonstrates 
that the following boundary conditions at $r=0$ must be imposed:
\begin{equation}
u_\uparrow^{v}(0)=c_v, \;\; \partial_r u_\uparrow^{v}(0)=u_\downarrow^{v}(0)=0, \;\; \partial_r u_\downarrow^{v}(0)=c_v'.
\label{eq:BC:origin}
\end{equation}
Here, we 
have labeled the functions $u_{\uparrow,\downarrow}$ with the 
superscript `$v$' to indicate a Majorana state localized near the vortex core. One of the unknown constants $c_v$ and $c_v'$ should be determined by the normalization condition, Eq.~\eqref{eq:norm:cond:m}, while the other should be  
chosen so that the Majorana state is localized at the origin, i.e., $u_{\uparrow,\downarrow}^{v}\to0$ at $r\to \infty$. 

In order to determine the 
condition for the existence of the Majorana state, we study the asymptotic behavior of the solution at $r\to \infty$. We can seek it in the following form, $\phi^\eta\sim \phi^{\eta}_{\infty}\exp(-Q r)$.
For $\phi^{\eta}_{\infty}$ to be nonzero, the wave vector $Q$ must satisfy the following characteristic equation (see Appendix~\ref{App:A}): 
\begin{equation}
\left (\frac{Q^2}{2m}+ \mu\right )^2+(\Delta_{\infty} - \eta \alpha Q)^2=J^2 .
 \label{t:eq:Q}
\end{equation}
This equation has four solutions, but only such of them contribute to the wave function $\phi^\eta$, which define decaying exponents at $r\to \infty$, i.e., solutions with $\re Q>0$. We emphasize that such $Q$ should be at least three, because $\phi^\eta$ %should 
contains three unknown constants. Indeed, in that case we have $2+3=5$ constants, $2$ from asymptotic behavior at $r\to 0$, see Eq.~\eqref{eq:BC:origin}, and $3$ from asymptotic behavior at $r\to \infty$, which are fixed by the normalization condition, cf. Eq.~\eqref{eq:norm:cond:m}, and by the four equations\footnote{Two equations from the continuity of $u_{\uparrow,\downarrow}^{v}$ and two more equations from the continuity of the derivative $\partial_ru_{\uparrow,\downarrow}^{v}$.} matching the two  
asymptotes at $r\to 0$ and $r\to \infty$ at the certain intermediate point. Oppositely, if two or less $Q$ has the positive real part, then the Majorana state does not exist.~\footnote{If Eq. \eqref{t:eq:Q} has only two solutions with $\re Q>0$ then the Majorana state might exist for the special fine-tuned choice of the parameters $\mu$, $J$, $\Delta_\infty$, and $\alpha$.}

Let us proceed 
by considering the case where the spin-orbit coupling is absent, $\alpha=0$. In this scenario, Eq.~\eqref{t:eq:Q} reduces to a biquadratic equation, whose solutions can found analytically. One can easily verify that 
there are less than three wave vectors $Q$ with a positive real part 
and, therefore, the Majorana state localized near the vortex core does not exist.

In the general case of a non-zero spin-orbit coupling, we analyze the roots of Eq.~\eqref{t:eq:Q} using the Routh-Hurwitz stability criterion and Routh's table method. As a result, we arrive at the following conclusion: the equation~\eqref{t:eq:Q} has three roots with positive real parts if and only if:
\begin{equation}
    \mu^2+\Delta_{\infty}^2 < J^2, \qquad \eta \alpha<0  .
    \label{eq:cond}
\end{equation}
Therefore, for 
$\alpha\neq0$
the zero mode of the Hamiltonian~\eqref{eq:3:HH} localized at the origin corresponds to 
${\eta=-\sgn \alpha}$. We note that a condition  similar to Eq.~\eqref{eq:cond} for the existence of the Majorana state in a superconductor 
in the presence of the skyrmion with several radial spin flips has been found in Ref. \cite{Yang2016}. In that work, the role of the spin-orbit coupling in Eq.~\eqref{eq:cond} was played by an inverse period of radial spin flips.

\subsubsection{The Majorana state localized at the edge}

Now, we will discuss the conditions for the existence of the Majorana state localized near the edge of the system. We assume that the system is in the form of a disk with a radius $L$, which is larger than all other relevant scales. 
This means that we set $A_\varphi(r)=0$, $\theta(r)=0$, and ${\Delta(r)=\Delta_\infty}$ near the edge.

To formulate the appropriate boundary conditions, we assume that the wave function is zero outside the system. 
Then, we impose the following boundary conditions:
\begin{equation}
{u}_{\uparrow}^{e}(L)={u}_{\downarrow}^{e}(L)=0, 
\quad 
\partial_r{u}_{\uparrow}^{e}(L)=c_e, \quad \partial_r{u}_{\downarrow}^{e}(L)=c_e'.
\label{eq:BC:Infty}
\end{equation}
Here we  
labeled the functions $u_{\uparrow,\downarrow}$ with  
superscript `$e$' to indicate the Majorana state localized near the system edge. One of the unknown constants $c_e$ and $c_e'$ should be determined by the normalization condition, cf. Eq.~\eqref{eq:norm:cond:m}, while the other should be 
chosen so that the Majorana state vanishes far from the system edge.

In order to construct the zero mode of the Hamiltonian~\eqref{eq:3:HH}, which is localized near the edge at $r=L$, we employ the results of the previous subsection, see Eq.~\eqref{t:eq:Q} and the speculation next to it. Then the Majorana state exists if Eq.~\eqref{t:eq:Q} has at least three solutions with a negative real part, $\re Q<0$, that occurs, similar to Eq.~\eqref{eq:cond}, if and only if:
\begin{equation}
    \mu^2+\Delta_{\infty}^2 < J^2, \qquad \eta \alpha>0  .
    \label{eq:cond:L}
\end{equation}
Therefore, in the case of 
$\alpha\neq0$
the Majorana state, which is localized near the edge, corresponds to  
${\eta=+\sgn \alpha}$.

The zero-mode solution $\phi^\eta$ for $r\sim L$ is then characterized by three unknown constants and evanesces with decrease of $r$. 
As shown in Appendix~\ref{App:B}, these three unknown constants, along with $c_{e}$ and $c_{e}'$, 
can be fixed by the boundary conditions~\eqref{eq:BC:Infty} and the normalization condition, Eq.~\eqref{eq:norm:cond:m}. 
Therefore, the inequality~\eqref{eq:cond:L} 
guarantees the existence of the zero mode of the Hamiltonian~\eqref{eq:3:HH}, which is localized near the  
rim of the disk.

In order to illustrate our general results, in the following sections we study analytically and numerically the Majorana state localized near the vortex core, see Section~\ref{Sec:Core}, and near the edge of the system, Section~\ref{Sec:Edge}. We highlight two specific cases of magnetic textures, 
no-skyrmion and coaxial skyrmion configurations, with superconducting vortex in a thin chiral ferromagnetic--superconductor heterostructure and show that both configurations can be treated in some general way, based on the local rotational transformation of the Hamiltonian, see subsection~\ref{sec:rotation}.  
Additionally, we analyze the bound states with non-zero angular momentum~$l$, see Section~\ref{Sec:finL}.  

\section{The Majorana state localized near the vortex core~\label{Sec:Core}}

Here we consider a thin chiral ferromagnet--superconductor heterostructure with
a superconducting vortex in a superconducting film. 
We also assume that in a thin ferromagnetic film
a skyrmion is either absent (see subsection~\ref{Sec:NoSk}) or present and is located coaxially with a vortex  (see subsection~\ref{Sec:Sk}).  
Below we 
use the following assumptions about the parameters of the heterostructure:
\begin{equation}
 d_S\sim d_F \ll \ell_w \ll \lambda=\lambda_L^2/d_S.   
 \label{eq:assump:11}
\end{equation}
The length scale $\ell_w$ controls a width of a domain wall in a ferromagnet.
Pearl length $\lambda=\lambda_L^2/d_S$ determines the length scale of the vortex magnetic field.

\subsection{No-skyrmion configuration\label{Sec:NoSk}}

Here we assume that the formation of a skyrmion is not energetically favorable. Nevertheless, the vortex magnetic field induces the non-homogeneous magnetic texture described by Eq.~\eqref{eq:MM} with the following magnetic angle \cite{Apostoloff2024}:
\begin{equation}
\theta(r)\simeq \theta_v(r) \equiv - 2 \gamma \ell_w \int_0^\infty dq \frac{q \mathcal{J}_1(q r)}{(2 q+1/\lambda)(1+q^2\ell_w^2)} .  
\label{eq:theta:b}
\end{equation}
Here, we assume $\ell_w\gg\xi_v$ and introduce the dimensionless 
 parameters ${\gamma= \zeta \ell_w/\lambda}$ and $\zeta=M_s\phi_0/(8\pi \mathcal{A})$
 with $\mathcal{A}$ to be the exchange stiffness constant in the ferromagnet. 
 We note that 
 $\zeta\approx 1\div10$ for typical ferromagnetic materials \cite{Katkov2024}.
 The result~\eqref{eq:theta:b} is derived under assumption $|\theta_v|\ll1$ that is consistent with experimentally reasonable condition $\gamma\lesssim 1$ \cite{Petrovic2021}. 
 The integral expression~\eqref{eq:theta:b} can be approximated as \cite{Katkov2024}
\begin{gather}
 \theta_v(r) \simeq 
-\gamma \begin{cases}
[r/(2\ell_w)]\ln (\ell_w/r) , & \qquad r\ll \ell_w ,\\
\ell_w/[r+r^2/(2\lambda)], & \qquad r\gg \ell_w .
\end{cases} 
\label{eq:theta:b:r}
\end{gather}
In the case of a thin superconducting film the vector potential created by a Pearl vortex, cf. Eq.~\eqref{eq:vortex:A},  can be approximated as 
\begin{equation}
    A_\varphi^{v}(r) \simeq 
  -\frac{\phi_0}{4\pi \lambda}  \begin{cases}
        (r/d_S)\ln(d_S/r), & \qquad r\ll d_S ,
      \\
        1/[1+r/(2\lambda)], & \qquad d_S\ll r . 
    \end{cases} 
    \label{eq:vortex:A:r}
\end{equation}
Using Eqs.~\eqref{eq:vortex:M},~\eqref{eq:theta:b}, and~\eqref{eq:theta:b:r}, we obtain
\begin{equation}
    A_\varphi^{m}(r) \simeq \frac{2\pi M_s d_F \theta_v(r)}{1+r/(2\lambda)} .
    \label{eq:MM:A:r}
\end{equation}
The relative magnitude of the two contributions to the vector potential, $A_\varphi^{v}$ and $A_\varphi^{m}$, is controlled by a dimensionless parameter $M_s^2 d_F\ell_w/ \mathcal{A}$. If this parameter is large, the vector potential will be dominated by the contribution from the induced magnetic texture. Conversely, if the parameter is small, the main contribution to $A_{\varphi}$ will come from the superconducting vortex. Typically, $\mathcal{A}\sim (M_s\ell_w)^2$ \cite{UFN-SK}, so $A^{v}_\varphi$ dominates over $A^{m}_\varphi$ due to the assumption $d_F\ll \ell_w$, cf. Eq.~\eqref{eq:assump:11}.

The analysis of the eigenfunctions of the Hamiltonian~\eqref{eq:3:HH} is complicated by the presence of a vector potential and a magnetic texture induced by a vortex. We note that the contribution to the vector potential from the vortex and the induced magnetic texture disappears in the formal limit of $\lambda\to\infty$. Therefore, it is useful to compare the Pearl length with a spin-orbit-induced length that determines the spatial extent of a Majorana state localized at a vortex. 

For the sake of concreteness, we assume that 
the spin-orbit coupling is relatively small:
\begin{equation}
  m \alpha^2\ll\dfrac{J^2-\Delta_\infty^2}{\Delta_\infty^2}
  \big(\sqrt{J^2-\Delta_\infty^2}-\mu\big). 
  \label{eq:Lambda_so1}
\end{equation}
Then by solving Eq.~\eqref{t:eq:Q} perturbatively in $\alpha$, we find that the Majorana state localized near the vortex core decays on the scale 
\begin{eqnarray}
\Lambda_{\rm so} = (J^2-\Delta_\infty^2)^{1/2}/(m|\alpha| \Delta_\infty).
\label{eq:Lambda_so}
\end{eqnarray}
We note that for $J-\Delta_\infty\sim J\sim \Delta_\infty$
 the length scale $\Lambda_{\rm so}$ is nothing but the standard length scale, 
$(m|\alpha|)^{-1}$,
induced by the spin-orbit coupling. For $J\gg\Delta_\infty$ the decay length is parametrically larger than a standard spin-orbit length, $\Lambda_{\rm so}\gg (m|\alpha|)^{-1}$ due to a large factor $J/\Delta_{\infty}$.

In the case when $\Lambda_{\rm so}\ll \lambda$, the Majorana state is located within the region where the magnetic field created by the superconducting vortex is concentrated. For ${r\ll \lambda}$ the corresponding vector potential $A_\varphi^{v}$ is much smaller than $\phi_0/r$ and can therefore be neglected in the Hamiltonian $H^{\pm}$, Eq.~\eqref{eq:3:HH}. Similarly,  $A_\varphi^{m}$ can also be ignored since it is usually smaller than $A_\varphi^{v}$  as discussed above. 
If the spin-orbit-induced length is greater than the Pearl length, $\Lambda_{\rm so}\gg \lambda$, we have not found any parameters that make the influence of $A_\varphi$ distinguishable in the spatial distribution of the Majorana state wavefunction. Therefore, we assume $A_\varphi=0$ for all practical purposes. Nevertheless, we present a corresponding analysis for $A_\varphi\neq0$ in Appendix~\ref{App:B}.

The non-uniform magnetization $\theta_v$ induced by the superconducting vortex is relatively small (in virtue of $\gamma\ll 1$) and can therefore be neglected at first glance.
Then, the problem for the Majorana state localized at the vortex reduces to the case of a uniform magnetization affecting quasiparticles in a superconductor film through the Zeeman effect. 
This simplified problem is solved approximately in the following subsections for two directions of the uniform magnetization, $\theta\equiv0$ and $\theta\equiv\pm\pi$. The case of small non-uniform magnetization, $\theta\neq0$, is solved in subsection~\ref{sec:theta_v}.

\subsubsection{Uniform magnetization $\theta\equiv0$}

Even for $A_\varphi=0$ and $\theta=0$ the analytical solution of Eq.~\eqref{eq:eq:M} is hardly possible. Assuming the exchange coupling~$J$ is the largest energy scale, 
\begin{equation}
{J+\mu \gg \max(\Delta_\infty, m \alpha^2)},
\label{eq:J_great}
\end{equation}
but the ratio $m\alpha^2 (J+\mu)/\Delta_\infty^2$ can be of an arbitrary magnitude,
we find solution of Eq.~\eqref{eq:eq:M} 
within the WKB-type approximation, 
that describes the Majorana state localized near the core of the superconducting vortex, 
\begin{align}
   u_{\uparrow}^{v}(r) & \simeq \tilde{c}_v e^{-\mathcal{S}(r)} 
   \big [
     \mathcal{J}_0(r/R_J)  -\big( R_J\Delta(r)/|\alpha| \big) \mathcal{J}_1(r/R_J)
   \big ] ,
   \notag \\ 
   u_{\downarrow}^{v}(r) & \simeq  - 2\tilde{c}_v(R_JJ/\alpha) e^{-\mathcal{S}(r)} \mathcal{J}_1(r/R_J) .
   \label{eq:JJ:up}
\end{align}
Here $\tilde{c}_v$ is a constant that should be found from the normalization condition, Eq.~\eqref{eq:norm:cond:m}, $\mathcal{J}_k(x)$ stands for the Bessel function of the first kind,
$R_J=1/\sqrt{2m(J+\mu)}$ is a characteristic length of `fast' oscillations, 
and 
\begin{equation}
    \mathcal{S}(r) = \frac{m |\alpha|}{J} \int_0^r dr_1 \Delta(r_1) 
    \label{eq:def:K}
\end{equation}
determines a `slow' decay of the Majorana state away from the vortex center.

\begin{figure*}[t]
    \includegraphics[width=0.95\textwidth]{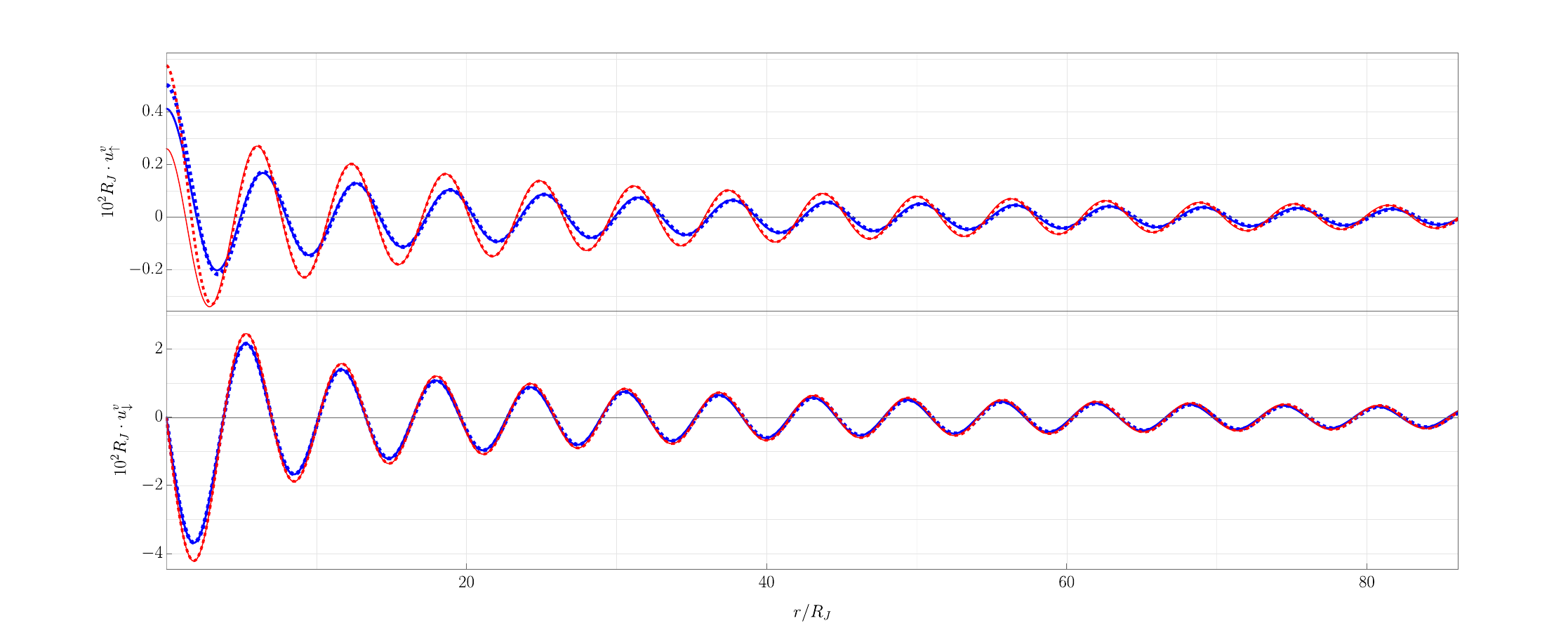}
    \caption{Majorana state localized at the vortex in the absence of a skyrmion: comparison of numerical and analytical solutions for $u_\uparrow^v$ (top panel) and $u_\downarrow^v$ (bottom panel). The solutions in the case of zero $\theta_v$ are presented with blue solid curves (numerics) and dashed curves (analytics, cf. Eq.~\eqref{eq:JJ:up}). The solutions in the case of non-zero $\theta_v$ (cf. Eq.~\eqref{eq:theta:b}) are presented with red solid curves (numerics) and dashed curves (analytics, cf. Eq.~\eqref{eq:JJ:up:after_rotation}). Numerical solutions are obtained with the help of the shooting method (see text). The wave functions are normalized according to Eq.~\eqref{eq:norm:cond:m}. We use the following parameters (in arbitrary units): $\mu=J/4$, $\Delta_\infty=J/8$, $(m\a)^{-1}=16R_J$, $\Lambda_{\rm so}=130R_J$, $\xi_v=1.5R_J$,    
    $\ell_w=30R_J$,
    $\l=360R_J$, and $\g=0.3$.}
    \label{fig:noSk}
\end{figure*}

We present the details of the derivation of Eq.~\eqref{eq:JJ:up} 
in Appendix~\ref{App:B}. The result~\eqref{eq:JJ:up} is valid formally at large distances, ${r\gg R_J}$, only. However, in practice, it works well for $r\gtrsim R_J$ that is demonstrated in Fig.~\ref{fig:noSk}, where we compare the result~\eqref{eq:JJ:up} with the numerical solutions of Eq.~\eqref{eq:eq:M} with $A_\varphi=\theta=0$.
The numerical solution 
shows excellent quantitative agreement with the analytical solution except near the origin.
Also, the solution~\eqref{eq:JJ:up} 
is 
consistent with the boundary conditions~\eqref{eq:BC:origin}, if one sets ${c}_v=\tilde{c}_v$ and ${c}_v'=-\tilde{c}_v J/\alpha$. 

The Bessel functions 
 $\mathcal{J}_0(r/R_J)$ and $\mathcal{J}_1(r/R_J)$ 
describe fast oscillations with the length scale $R_J$, while the function $\mathcal{S}(r)$ describes the slow decay of the oscillation amplitude with the length scale $\Lambda_{\rm so}$. 
As we said, our analytical solution works well provided $\D_\infty\ll J+\mu$. Increasing the value of $\Delta_\infty$ enhances the Majorana's localization and changes oscillations' length scale (analytical expression for this change is derived in Appendix~\ref{App:B}, see Eq. \eqref{eq:W(r)}). The localization of the Majorana state depends also on vortex radius $\xi_v$: the solution becomes less localized as $\xi_v$ increases.
When $\Delta_\infty$ reaches $\sqrt{J^2-\mu^2}$, the localization becomes non-exponential and the Majorana state disappears, cf. inequality~\eqref{eq:cond}.

%%%%%%%%%%%%%%%%%%%%%%%%%%%%%%%%%%%%%%%%%%%%%%%%%%%%%%%%%%%

%%%%%%%%%%%%%%%%%%%%%%%%%%%%%%%%%%%%%%%%%%%%%%%%%%%%%%%%%%%%

\subsubsection{Uniform magnetization $\theta\equiv\pm\pi$\label{sec:theta_pi}}

For the most part of the paper we assume that magnetization~$\bm{m}$ is directed in the $z$-direction far from the vortex, $\bm{m}\to\bm{e}_z$ at $r\to\infty$. However, it is instructive to discuss shortly the case with homogeneous $\bm{m}=-\bm{e}_z$, or $\theta(r)\equiv \pm \pi$. 

The speculations as given in the beginning of the subsection~\ref{Sec:NoSk} hold true here. Again we consider Eq.~\eqref{eq:eq:M} with the boundary conditions~\eqref{eq:BC:origin}, but apply  $\theta(r)\equiv \pm \pi$ and $A_\varphi=0$.
Then, analogously to Eq.~\eqref{eq:JJ:up}, we find solution within the WKB-type approximation (within condition~\eqref{eq:J_great}), 
\begin{align}
   u_{\downarrow}^{v,\pi}(r) & \simeq  \tilde{c}_v e^{-\mathcal{S}(r)} 
   \big [
     \mathcal{J}_1(r/R_J)  +\big( R_J\Delta(r)/|\alpha| \big) \mathcal{J}_0(r/R_J)
   \big ], \notag \\
   u_{\uparrow}^{v,\pi}(r) & \simeq - 2\tilde{c}_v(R_JJ/\alpha) e^{-\mathcal{S}(r)} \mathcal{J}_0(r/R_J) ,
   \label{eq:JJ:up:pi} 
\end{align}
where $\tilde{c}_v$ is a constant that should be found from the normalization condition, Eq.~\eqref{eq:norm:cond:m}, and $\mathcal{S}(r)$ is defined in Eq.~\eqref{eq:def:K}. The details of the derivation of Eq.~\eqref{eq:JJ:up:pi} can be found in Appendix~\ref{App:B}.

\subsubsection{Rotation for the non-uniform magnetization\label{sec:rotation}}

In order to solve the eigenvalue problem~\eqref{eq:eq:M} in the presence of spatially dependent angle $\theta$, it is convenient to perform a unitary transformation of the Hamiltonian $H^{\pm}$ and rotate the solution $\phi^{\pm}(r)$ as
\begin{equation}
	\underline{H}^\pm  = e^{i\vartheta\sigma_y/2} H^\pm e^{-i\vartheta\sigma_y/2},
	\quad 
	\phi^{\pm}=e^{-i\vartheta\sigma_y/2} \underline{\phi}^\pm,
	\label{eq:eq:M:R}
\end{equation}
where function $\vartheta(r)$ can be conveniently chosen as 
 $\vartheta(r)=\theta(r)-\theta(r=0)$. 
Since $\vartheta(r=0)=0$, no rotation is needed for the boundary condition at $r=0$. 

Due to radial symmetry of the system ${\theta(r=0)=\chi \pi}$, where $\chi$ is an integer number. Particularly, $\chi=0$ for the no-skyrmion configuration and $\chi=\pm1$ for the skyrmion.
Then Eq.~\eqref{eq:eq:M} becomes $\underline{H}^\pm\, \underline{\phi}^\pm = 0$ 
where the transformed Hamiltonian is given explicitly as 
\begin{widetext}
	\begin{align}
		\underline{H}^\eta  = & -\frac{1}{2m}\Bigl (\partial_r^2 +\frac{1}{r}\partial_r\Bigr ) 
		+\frac{1}{2m} \Bigl (\frac{\sigma_z\cos\vartheta  -\sigma_x \sin \vartheta-1}{2r}+e A_{\varphi}\Bigr )^2 
		-\mu + i  \Bigl (\eta\Delta+\frac{\partial_r^2\vartheta}{4m}\Bigr ) \sigma_y  +(-1)^\chi J \sigma_z  
		\notag \\
		& +\frac{\alpha \partial_r \vartheta}{2} 
        +\frac{(\partial_r \vartheta)^2}{8m} 
        + i \Bigl (\alpha+\frac{\partial_r\vartheta}{2m}\Bigr ) 
            \sigma_y \left (\partial_r+\frac{1}{2r}\right ) 
		+\alpha \Bigl (\frac{1}{2r}-e A_{\varphi}\Bigr ) 
            \Bigl (\sigma_x \cos\vartheta +\sigma_z \sin\vartheta\Bigr )  .
		\label{eq:3:HH:R}
	\end{align}
\end{widetext}

Then the functions $u_{\uparrow,\downarrow}(r)$ is related to $\underline{u}_{\uparrow,\downarrow}(r)$ as
\begin{equation}
	\begin{split}
		u_{\uparrow}(r) & = \underline{u}_{{\uparrow}}(r) \cos\frac{\vartheta(r)}{2} -\underline{u}_{{\downarrow}}(r) \sin \frac{\vartheta(r)}{2} ,  \\
		u_{\downarrow}(r) & = \underline{u}_{{\uparrow}}(r) \sin\frac{\vartheta(r)}{2} +\underline{u}_{{\downarrow}}(r) \cos \frac{\vartheta(r)}{2} .
	\end{split}
	\label{eq:MM:sol:rotate}
\end{equation} 

\subsubsection{Analytical solution for small non-uniform $\theta(r)$\label{sec:theta_v}}

Now we can calculate an approximate solution for the case of non-uniform magnetization, 
$\vartheta(r)=\theta(r)=\theta_v(r)$,
which is produced by the superconducting vortex, see Eq.~\eqref{eq:theta:b}. Assuming 
\begin{equation}
    \lambda\gg\ell_w\gtrsim R_J \gtrsim \max\left [\left (\frac{\gamma \ell_w}{m|\alpha|}\right )^{1/2}, 
    \left(\frac{\gamma \ell_w}{m \Delta}\right )^{1/3}\right ]
\end{equation}
\color{black}
we can neglect the angle $\theta(r)$ and its derivatives in the Hamiltonian~\eqref{eq:3:HH:R}.

Then the solution of Eq.~\eqref{eq:eq:M} can be written as Eq.~\eqref{eq:MM:sol:rotate} where $\underline{u}_{\uparrow,\downarrow}$ should be taken as the solution for $\theta\equiv0$ from Eq.~\eqref{eq:JJ:up}. As a result, we obtain
\begin{align}
   u_{\uparrow}^{v}(r)  \simeq {}&
   \tilde{c}_v e^{-\mathcal{S}(r)} 
   \Big [
     \mathcal{J}_0(r/R_J)  
     \notag\\
     &{\qquad}-\big\{ R_J[\Delta(r)-J\theta_v(r)]/|\alpha| \big\} \mathcal{J}_1(r/R_J)
   \Big ] ,
   \notag \\ 
   u_{\downarrow}^{v}(r)  \simeq  {}&
   \tilde{c}_v e^{-\mathcal{S}(r)} 
   \Big \{
     [\theta_v(r)/2]\mathcal{J}_0(r/R_J)  
     \notag\\
     &{\qquad}-2\big(R_JJ/|\alpha| \big) \mathcal{J}_1(r/R_J)
   \Big \}.
   \label{eq:JJ:up:after_rotation}
\end{align}
Figure~\ref{fig:noSk} demonstrates that non-zero $\theta_v$ can affect the solution, especially $u_{\uparrow}^{v}$, see red solid (numerics) and dotted (analytics, cf. Eq.~\eqref{eq:JJ:up:after_rotation}) curves, in comparison with the case of $\theta_v=0$. This happens because $\Delta(r)$ and $J\theta_v(r)$ can 
to be comparable in magnitude if $\gamma\sim\Delta_{\infty}/J$. 
We note that parameters $J$, $\alpha$, and $\xi_v$ employed in our numerical calculations are similar to those for a heterostructure studied experimentally in Ref.~\cite{Petrovic2021} provided ${\Delta_{\infty}=1.5}$~meV, ${R_J=2}$~nm, and $m$ is taken as the electron mass.

\begin{figure*}[t]
    \includegraphics[width=\textwidth]{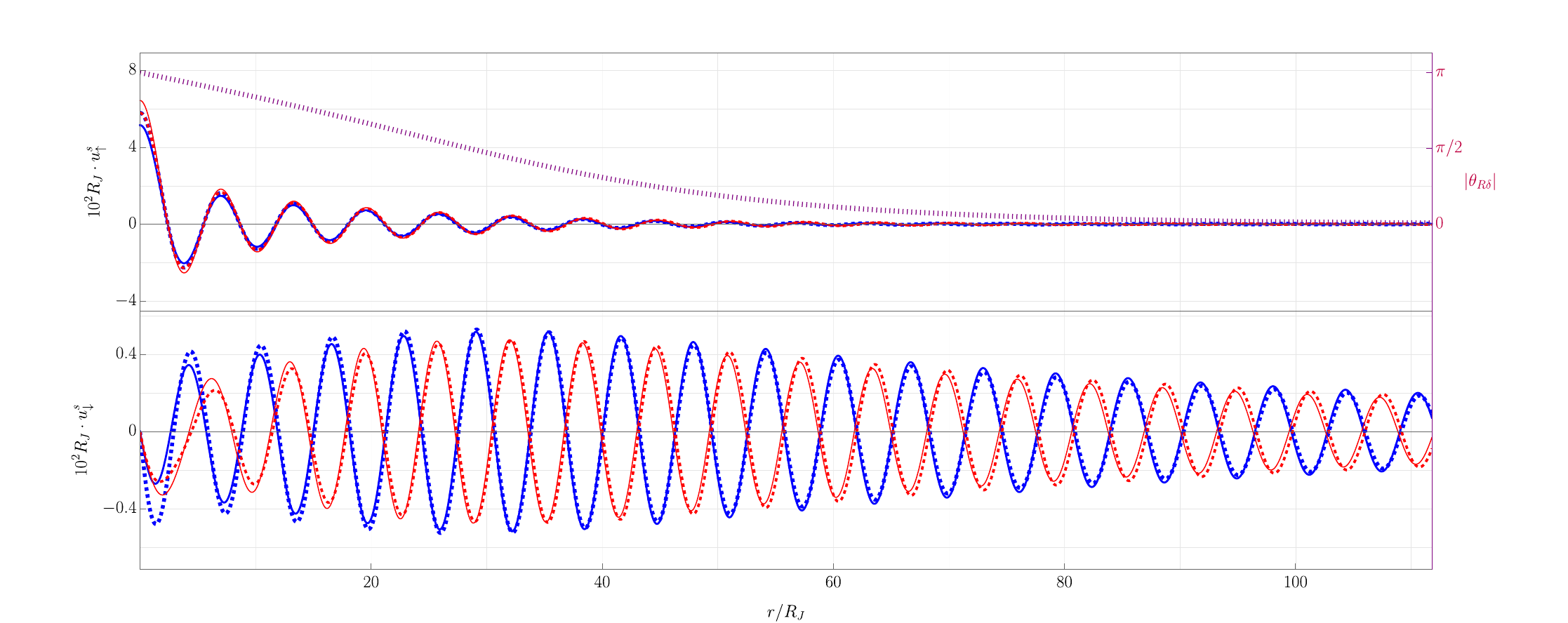}
	\caption{Majorana state localized at the vortex in the presence of a skyrmion: comparison of numerical and analytical solutions. Solid curves are numerics, dashed curves are analytics, cf. Eq.~\eqref{eq:MM:sol:Sk}; blue curves corresponds to $\chi=1$ and red ones to $\chi=-1$. Also the profile $\theta_{R\delta}$, given by Eq.~\eqref{eq:sk:theta}, is plotted in purple. We use $|R|=30R_J,\,\delta=20R_J$, and the other parameters coincide with those in Fig. \ref{fig:noSk}.}
	\label{fig:Sk}
\end{figure*}

\subsection{Coaxial skyrmion\label{Sec:Sk}}

In this subsection, we will focus on the situation where there is a skyrmion situated coaxially with a superconducting vortex. As shown in Ref.~\cite{Apostoloff2023}, the exact profile of the skyrmion centered at the vortex can be effectively approximated by the following coaxial ansatz:
\begin{gather}
	\theta(r)\simeq \theta_{R\delta}^{\gamma}(r) \equiv  \theta_{R\delta}(r) + \theta_v(r)\cos \theta_{R\delta}(r) .
	\label{eq:coax:an}
\end{gather}
Here, $\theta_v(r)$ is given by Eq.~\eqref{eq:theta:b} and takes into account the perturbation of the magnetization by the magnetic field generated by the vortex. The function $\theta_{R\delta}(r)$ describes the 360$^\circ$ domain wall (DW) ansatz:
\begin{gather}
	\theta_{R\delta}(r)= 2 \arctan \frac{\sinh(R/\delta)}{\sinh(r/\delta)} ,
	\label{eq:sk:theta}    
\end{gather}
where $\delta$ is the effective domain wall width and the parameter $R$ represents the skyrmion radius ($|R|$) and the chirality $\chi=\sgn R$. Since $\theta_{R\delta}^{\gamma}(0)=\pi \chi$, the skyrmion angle is not small, in contrast to $\theta_v$. 
Note that skyrmions of both chiralities $\chi=\pm 1$ can be stabilized due to the stray fields of the superconducting vortex. 
As our numerical results have shown, $\theta_v$ in Eq.~\eqref{eq:coax:an} can be neglected 
for the studying of the Majorana states.
However, accounting for $\theta_v$ is necessary to determine the correct parameters $R$ and $\delta$ when the micromagnetic problem should be solved~\cite{Apostoloff2023}. In realistic systems the radius $R$ and wall width $\delta$ are of order of the magnetic length~$\ell_w$. Moreover, the magnetic field of the superconducting vortex forces the radius of the skyrmion to enlarge~\cite{Xie2023}, therefore, it can be assumed that $R\gtrsim\delta\sim\ell_w$.

The vector potential generated by the superconducting vortex can still be described by Eq.~\eqref{eq:vortex:A:r}. The magnetic texture created by the skyrmion produces the vector potential that cannot be easily expressed in term of the skyrmion angle $\theta(r)$, see Eq.~\eqref{eq:vortex:M}. \color{black} Typically, for $\lambda\gg R$ $A^{m}_\varphi$ is non-monotonic function of distance with the maximal magnitude of order of $M_s d_F$ at distance $r\sim R$ from the skyrmion's center and fast decaying for $r\gg\lambda$ \cite{Andriyakhina2021}. Thus we can estimate the ratio of maximal magnitudes of $A^{m}_\varphi$ and $A^{v}_\varphi$ as $M_sd_F\lambda/\phi_0$.  
Thus, this dimensionless parameter controls the importance of $A^{m}_\varphi$ in comparison with $A^{v}_\varphi$. For $M_sd_F\lambda/\phi_0\ll 1$ one can neglect the vector potential induced by the skyrmion in comparison with the vector potential induced by the vortex. The condition $M_sd_F\lambda/\phi_0\sim 1$ indicates that both contributions to the vector potential are equally important for distances $r\lesssim\lambda$ at least.  
\color{black}

\begin{figure*}[t]
    \includegraphics[width=\textwidth]{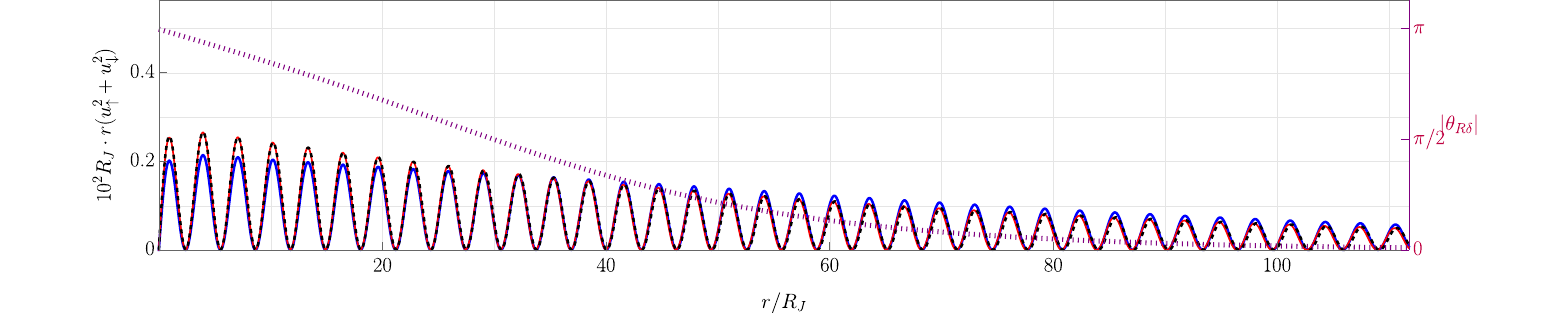}
	\caption{Radial probability density of the Majorana state localized at a vortex. Numerical results are presented for homogeneous magnetization $\theta\equiv\pi$ (red) and skyrmion $\theta=\theta_{R\delta}$ (blue). Black dashed curve is  analytical solution in accordance with Eq.~\eqref{eq:MM:sol:Sk}. 
    Also the profile $\theta_{R\delta}$, given by Eq.~\eqref{eq:sk:theta}, is plotted in purple. For skyrmion we use $|R|=30R_J,\,\delta=20R_J$, and the other parameters coincide with those in Fig. \ref{fig:noSk}.}
	\label{fig:Sk:density}
\end{figure*}

\subsubsection{Analytical expression}

In order to derive an analytical expression for the Majorana state we use the rotation of the Hamiltonian, described in subsection~\ref{sec:rotation}. 
It seems impossible to treat the Hamiltonian~\eqref{eq:3:HH:R} analytically in its full glory. In order to demonstrate the effect of a skyrmion on the Majorana state localized near the center of the vortex,  
we make the same suppositions~\eqref{eq:J_great} as in Sec.~\ref{Sec:NoSk}, i.e.,
\begin{equation}
\min[(m\Delta_\infty)^{-1/2}, (m \alpha)^{-1}]\gg R_J .   
\label{eq:Rdelta_cond:0}
\end{equation}
Additionally, we assume that the skyrmion angle changes slowly on a length scales induced by the spin-orbit coupling and the superconducting gap,
\begin{equation}
\min[R,\delta]\gg \max[(m\Delta_\infty)^{-1/2}, (m \alpha)^{-1}] ,
\label{eq:Rdelta_cond}
\end{equation}
and thickness $d_F$ of the ferromagnet film is small enough, 
$M_s \lambda d_F/\phi_0\ll 1$.
Under these assumptions, we can neglect spatial derivatives of $\vartheta$ and $A_\varphi$ in Eq.~\eqref{eq:3:HH:R}, as well as 
set $\vartheta(r)=0$ in the kinetic and spin-orbit terms.

Therefore, the solutions~$\underline{u}_{\uparrow,\downarrow}$ are approximated by the functions~${u}_{\uparrow,\downarrow}^{v,\pi}$ for the homogeneous case with $\theta=\pi$.
Rotating it back by the angle $\vartheta(r)=\theta(r)-\chi \pi$ in accordance with Eqs.~\eqref{eq:MM:sol:rotate} we arrive at
\begin{equation}
	\begin{split}
		u_{\uparrow}^s(r) & = \chi\Big[{u}_{{\downarrow}}^{v,\pi}(r) \cos\frac{\theta(r)}{2} +{u}_{{\uparrow}}^{v,\pi}(r) \sin \frac{\theta(r)}{2}
        \Big],  \\
		u_{\downarrow}^s(r) & = \chi\Big[{u}_{{\downarrow}}^{v,\pi}(r) \sin\frac{\theta(r)}{2} -{u}_{{\uparrow}}^{v,\pi}(r) \cos \frac{\theta(r)}{2}
        \Big].
	\end{split}
    \label{eq:MM:sol:Sk}
\end{equation} 
Here we use the indices ``s'' instead of ``v'' in order to indicate the presence of the skyrmion. 
Physically, this result means that the Majorana state localized near the vortex core is determined primarily by the magnetization of the center of skyrmion ($\theta=\chi\pi$), but then straightforwardly should be rotated due to the changing of the skyrmion angle.

The details of the calculation leading to the results~\eqref{eq:MM:sol:Sk} are given in Appendix~\ref{App:B}. It should be noted that the functions $u_{\uparrow,\downarrow}^s$ are actually more complicated than given by Eq.~\eqref{eq:MM:sol:Sk}. Nevertheless, Eq.~\eqref{eq:MM:sol:Sk} correctly describes the behavior of the Majorana wave function quantitatively for $r\gg R_J$ and qualitatively for $r\lesssim R_J$ (see Fig.~\ref{fig:Sk}). Interestingly, in the considered assumptions, 
the decay rate is independent of the non-zero skyrmion angle and is solely determined 
by the spin-orbit coupling and the superconducting gap. 
The skyrmion angle produces small correction to the decay rate, as well as to the oscillation frequency; the details can be found in Appendix~\ref{App:B}, see Eqs. \eqref{eq:Svartheta}, \eqref{app:eq:Auu:113}.

Figure~\ref{fig:Sk} shows the components~$u_{\uparrow,\downarrow}^s$ of the Majorana state as the function of a distance from the vortex core. At $r\ll R$, the upper component~$u_\uparrow$ is prevalent over $u_\downarrow$ in amplitude because $\theta(r)\approx\pi$. At $r\gg R$, the magnetization is opposite, $\theta(r)\approx0$, and $u_{\uparrow}$ becomes negligible in comparison with $u_\downarrow$, which resembles its behavior for the homogeneous magnetization (cf. Fig.~\ref{fig:noSk}).

Figure~\ref{fig:Sk:density} illustrates another feature of the solution \eqref{eq:MM:sol:Sk}. Since this solution is obtained from ${u}_{\uparrow,\downarrow}^{v,\pi}$ by a rotation, the probability density of the Majorana state $(u_\ua^s)^2+(u_\da^s)^2$ in the presence of the skyrmion coincides with that for the case of homogeneous magnetization $\theta\equiv\pi$.

\begin{figure*}[t!]
	\includegraphics[width=\textwidth]{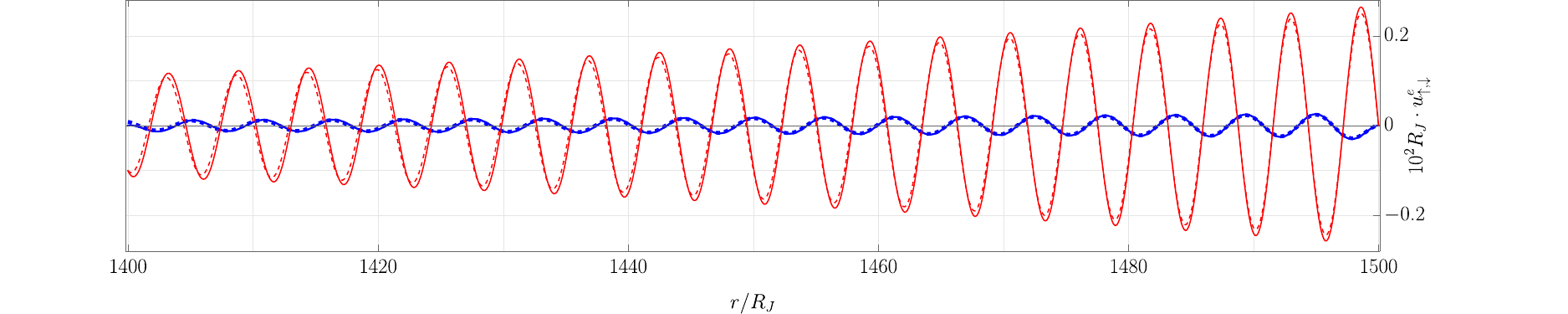}
	\caption{Majorana state localized at the edge: comparison of numerical and analytical solutions. The component $u_\uparrow$ is shown in blue solid (numerics) and dashed (analytics), while the component $u_\downarrow$ is shown in red solid (numerics) and dashed (analytics). Analytical solutions are given by Eq.~\eqref{eq:JJ:up:edge}. We use the same parameters as in Fig. \ref{fig:noSk} with the system size $L=1500R_J$.}
	\label{fig:Skr}
\end{figure*}

\subsubsection{Survival probability}

As we have seen above, the Majorana state is located at the center of the vortex, whether there is a skyrmion presents or not. However, because the skyrmion can be disrupted by fluctuations, it is important to estimate the probability that the Majorana state will still be located at the vortex center after the skyrmion has vanished. Using the solutions~\eqref{eq:JJ:up} and~\eqref{eq:MM:sol:Sk}, computing the overlap of the two wave functions, 
and assuming that skyrmion is not very large, $R\ll \Lambda_{\rm so}$, we estimate the probability to stay in the Majorana state as
\begin{equation}
    p \simeq \bigg ( \frac{2m \alpha^2 \Delta_\infty}{J^2R_J} \int\limits_0^\infty  dr\, \sin [\theta(r)/2]\bigg )^2.
\end{equation}
In the case of $R\gtrsim\delta$, we can calculate the last integral approximately using the DW ansatz, cf. Eq.~\eqref{eq:sk:theta},
\begin{equation}
	p \simeq  \frac{8m \alpha^2(J+\mu)}{J^2}\left (\dfrac{R+\delta\ln2}{\Lambda_{\rm so}}\right )^2  .
    \label{eq:prob}
\end{equation}
Therefore, we see that the probability of the Majorana state to survive after the skyrmion was destroyed is small. \color{black} 
For parameters $R = 30R_J$ and $\delta = 20R_J$, estimate \eqref{eq:prob} yields a probability $p \simeq 0.012$, while numerical computation using numerically exact wave functions gives $p \simeq 0.016$.

\section{The Majorana state localized near the edge\label{Sec:Edge}} 

In this Section we study the Majorana state localized near the edge of the system.
We make the same assumptions about the parameters of the system as in  
Sec.~\ref{Sec:Core} and assume that the system is in the form of a disk with a radius $L$. For the Majorana state localized near the rim, %edge, 
one has to compare~$L$ and the spin-orbit length $\Lambda_{\rm so}$, see Eq.~\eqref{eq:Lambda_so}. For the large enough system radius, $L\gg \Lambda_{\rm so}$, one can omit $A_\varphi(r)$ and $\theta(r)$ from the Hamiltonian~\eqref{eq:3:HH}. In this case, the Majorana state localized near the edge does not depend on the magnetic configuration at the origin (skyrmion or not) and can be found analytically. 
This conclusion is fully supported by the data shown in Fig.~\ref{fig:Skr}. 

The solution~$\phi^\eta(r)$ can be expressed as a linear combination of three exponents corresponding to three roots of Eq.~\eqref{t:eq:Q} with $\re Q<0$, 
\begin{align}
    u_{\uparrow}^{e}(r) &= c_1 e^{-Q_1  r}+c_2 e^{-Q_2  r}+c_3 e^{-Q_3  r},
    \notag \\
    u_{\downarrow}^{e}(r) &= c_1 \zeta_{Q_1} e^{-Q_1  r}+c_2 \zeta_{Q_2} e^{-Q_2  r}+c_3 \zeta_{Q_3} e^{-Q_3  r}.
    \label{eq:App:A:pp:an:1}
\end{align}

The exact expressions for $c_{j}$, $Q_j$, and $\zeta_Q$ are given in Appendix~\ref{App:C}. Here we present the asymptotic expressions for $u_{\uparrow,\downarrow}^{e}(r)$ assuming that the exchange coupling~$J$ is the largest energy scale, see Eq.~\eqref{eq:J_great},
\begin{align}
   & u_{\uparrow}^{e}(r)  \simeq  
     c_e e^{-\tilde{r}/\tilde{R}_J}
   \notag \\ 
     & \quad{}
     -c_e e^{-\tilde{r}/\Lambda_{\rm so}}
     \big\{
       \cos(\tilde{r}/R_J)
       -\big( R_J\Delta_{\infty}/|\alpha| \big) \sin(\tilde{r}/R_J)
   \big \} ,
   \notag \\ 
   & u_{\downarrow}^{e}(r)  \simeq - 2c_e(R_JJ/\alpha) e^{-\tilde{r}/\Lambda_{\rm so} } \sin(\tilde{r}/R_J).
   \label{eq:JJ:up:edge}
\end{align}
Here $\tilde{r}=L-r$ is the radial distance from the rim of the disk, %edge, 
$\Lambda_{\rm so} \simeq J/(m|\alpha|  \Delta_{\infty})$, and we introduce the length scale $\tilde{R}_J=1/\sqrt{2m(J-\mu)}$. The constant $c_e$ should be found from the normalization condition~\eqref{eq:norm:cond:m}, namely, $c_e\simeq\alpha/(2JR_J \sqrt{\pi L \Lambda_{\rm so}})$.

Figure~\ref{fig:Skr} shows the good agreement between the analytics,  Eq.~\eqref{eq:JJ:up:edge}, and numerics for the Majorana state localized at the edge.

\begin{figure}[t!]
	\includegraphics[width=0.95\columnwidth]{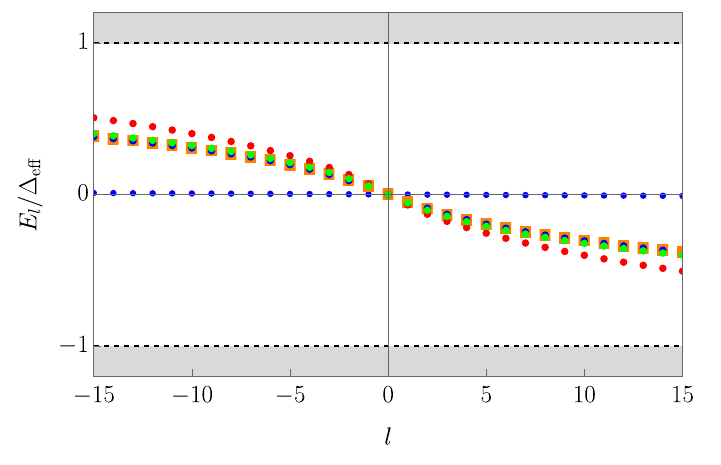}
	\caption{The spectrum of the localized states as a function of the angular momentum $l$ in the absence of a skyrmion.  
The energies $E_l^v$ for for the case $\theta=0$ are shown by blue (numerics), green (Eq.~\eqref{eq:El:WKB}), and red (Eq.~\eqref{eq:El:main}) points. 
Orange rectangles stand for the numerical results for energies $E_l^v$ in the case $\theta=\pi$.  We note that orange rectangles almost coincide with blue points, in accordance with Eq.~\eqref{eq:El:WKB:Sk:R}. The blue point near zero are the energies of states localized et the edge (see Fig.~\ref{fig:spedge}). Black dashed lines depict the energy gap level, cf. Eq. \eqref{eq:Delta:eff}. The gray bars show parts of the spectrum where all states are delocalized. All parameters are the same as in Fig.~\ref{fig:noSk}. 
        }
	\label{fig:sp}
\end{figure}

\section{Bound states with non-zero angular momentum\label{Sec:finL}}

The system spectrum contains both bound and delocalized states. The latter are separated from Majorana states with the effective energy gap $\D_{\rm eff}$. To estimate $\D_{\rm eff}$ we consider the Hamiltonian \eqref{eq:Hl} at sufficiently large~$r$, where $\theta=0,\,A_\varphi=0,\,\D=\D_\infty$, and all terms of order of $1/r$ and $1/r^2$ can be neglected,
\begin{gather}
H^{(l)}= -\frac{\tau_z}{2m}\partial_r^2 -\mu \tau_z +\Delta_\infty \tau_x  +J \sigma_z + i \alpha\partial_r\tau_z\sigma_y. 
\label{eq:Hasymp} 
\end{gather}
We note the above Hamiltonian coincides with the Hamiltonian for a 1D wire with spin-orbit coupling $\alpha$, Zeeman splitting $J$, and the superconducting gap $\Delta_\infty$ \cite{Oreg2010}.   
Then we solve Eq.~\eqref{eq:5.0} assuming eigenstate ${\Phi_l\propto\exp(-ipr)}$ and find two branches of the continuous spectrum,
\begin{gather}
   E_{\pm}^2=J^2+\xi_p^2+\Delta^2+\a^2p^2\pm2\sqrt{J^2(\xi_p^2+\Delta^2)+\xi_p^2\a^2p^2},
\end{gather}
where $\xi_p=p^2/(2m)-\mu$. Minimizing $|E_{-}|$ over $p$ within the assumption $\max(m\alpha^2, \Delta) \ll J$, cf. Eq. \eqref{eq:J_great}, we find the effective energy gap,
\begin{gather}
\D_{\rm eff} \simeq \frac{\a\Delta_\infty}{J R_J}
\ll\Delta_\infty .
\label{eq:Delta:eff}
\end{gather}
We note that $\D_{\rm eff}$ can be interpreted as the gap for `p-wave' superconducting state induced in the presence of a strong Zeeman splitting.  

Further we analyze the spectrum of the bound states with energy below the effective gap by two methods, numeric and analytic. \color{black} The numerical method is based on a finite-difference scheme for the 1D radial problem on a uniform grid with $N$ points. Let us explain it in more detail. Instead of using $\Phi_l$ from Eq.~\eqref{eq:5.0}, the method works with $\chi_l(r) = \sqrt{r}\,\Phi_l(r)$, for which zero boundary conditions are imposed, i.e., $\chi_l(0) = \chi_l(L) = 0$. All derivatives are treated using fourth-order approximations. For the resulting $4N \times 4N$ matrix, several eigenvalues (and corresponding wave functions) near $E = 0$ are obtained. In our calculations, we use $L = 1500R_J$ and $N = 8000$.

The analytical method is described in Appendix~\ref{App:D} and gives us the approximate expression for the bound states $\Phi_l$. Then we can  
estimate the energies straightforwardly as $E_l=\langle\Phi_l|H^{(l)}|\Phi_l\rangle$, where $H^{(l)}$ is given in Eq.~\eqref{eq:Hl}.

\subsection{No-skyrmion configuration}

We start with the bound states in the absence of a skyrmion. 
Figure~\ref{fig:sp} shows the energies $E_l$ of the bound states for different angular momentum~$l$. Two spectral branches are distinguished here. The first branch (with lower energies) corresponds to the states localized at the edge of the system and has a linear dispersion. The second branch describes the energies of the states localized near the vortex and has more complicated  
dependence of $l$. 
Below, we discuss the localized states corresponding to these two spectral branches in more detail.

\begin{figure}[t!]
	\includegraphics[width=0.95\columnwidth]{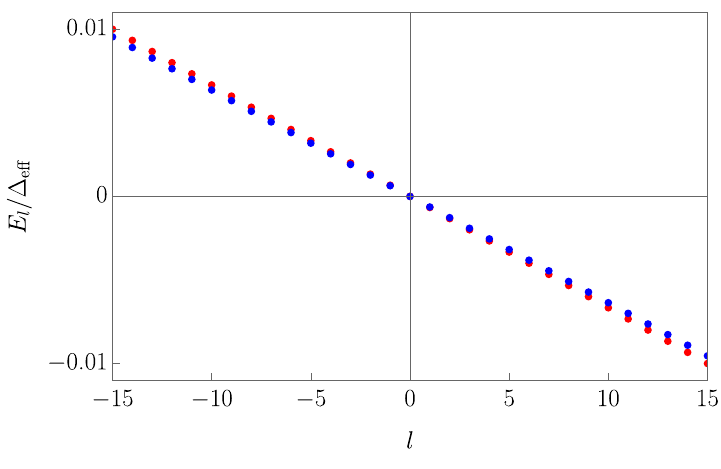}
	\caption{Comparison of numerical (blue) and analytical (red),  cf. Eq. \eqref{eq:El:LLL}, dependence of the energies of the states localized at the edge on the angular momentum $l$.     
    All parameters are the same as in Fig. \ref{fig:noSk}.}
	\label{fig:spedge}
\end{figure}

\subsubsection{Spectrum of the edge-localized states}

The 
spectrum of the states localized near the edge of the system can be easily described perturbatively. Indeed, let's write the Hamiltonian ~\eqref{eq:Hl} as ${H^{(l)}\simeq H^{(0)}+V_l}$, where 
	\begin{equation}
		V_l=\dfrac{l(1-\sigma_z\tau_z)+l^2\tau_z}{2m r^2}+\dfrac{\alpha l \sigma_x\tau_z}{r},
	\end{equation}
Here, we neglect the effect of the vector potential as in the previous sections.

At large $r$, which are only important for the  
considered spectral branch, $V_l$ is relatively small. Then, energies $E_l$ of corresponding states can be found as
\begin{equation}
	E_l  \simeq \langle \Phi_0^\eta | V_l | \Phi_0^\eta \rangle =  
4\pi l \int\limits_0^\infty dr \Bigl [\frac{u_{\downarrow}(r)}{2mr}
	+\alpha u_{\uparrow}(r)  \Bigr ]  u_{\downarrow}(r) .
	\label{eq:El:app:0}
\end{equation}
Here $\eta=\sgn(\a)$, so that $\Phi_0^\eta$ is the Majorana state localized near the edge and the functions $u_{\uparrow,\downarrow}(r)$ are normalized as in Eq.~\eqref{eq:norm:cond:m}.
Assuming that the system size $L$ is the longest lengthscale, we obtain the following result from Eq.~\eqref{eq:El:app:0}: 
\begin{gather}
	E_l \simeq \frac{2 |\alpha| l}{L} \frac{I_{01}}{I_{00}+I_{11}} 
	\simeq  - \frac{|\alpha| \Delta_\infty l}{J L} .
	\label{eq:El:LLL}
\end{gather}
Here, the quantities $I_{ab}$ are defined in Eq.~\eqref{eq:III:def} in Appendix~\ref{App:C}. 
A simplified expression is derived under  
condition~\eqref{eq:J_great} and can be directly calculated by using solution from Eq.~\eqref{eq:JJ:up:edge}. The slope of $E_l$ is negative for the states localized near the edge, $dE_l/dl<0$ independent of the sign of the spin-orbit coupling $\alpha$. The result \eqref{eq:El:LLL} is in perfect agreement with numerical findings.

We note that Eq.~\eqref{eq:El:LLL} allows us to interpret the states localized at the edge of the system as a continuous chiral state with a linear dispersion and velocity $-\Delta_{\rm eff} R_J$ propagating along the rim.

\begin{figure}[t!]
	\centerline{\includegraphics[width=0.95\columnwidth]{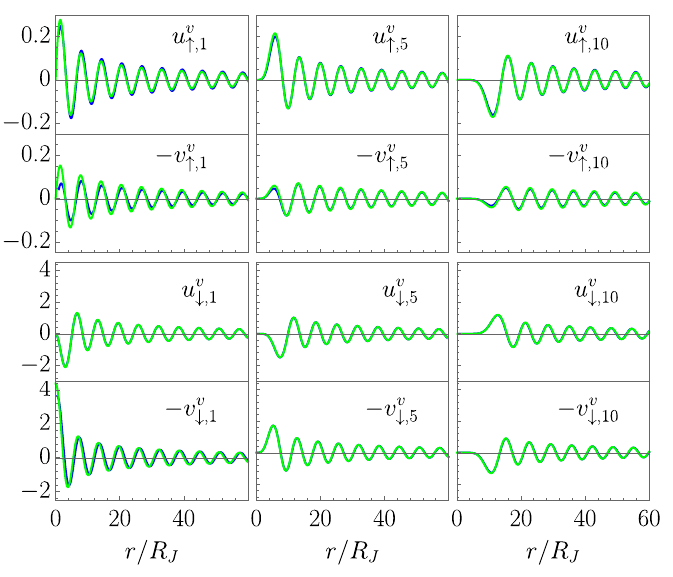}}
	\caption{Wave functions for the states localized at a vortex in the absence of a skyrmion: comparison of numerical (blue) and analytical (green) solutions for     $l=1,\,5,$ and $10$. Analytical solutions are given by Eqs. \eqref{eq:up:WKB:lnot0}, and \eqref{eq:down:WKB:lnot0}. All parameters are the same as in Fig. \ref{fig:noSk}. Wave functions are normalized according to Eq. \eqref{eq:norm:cond:m} and multiplied by a factor $10^2R_J$.}
	\label{fig:lnonzero}
\end{figure}

\begin{figure}[!htbp]
	\centerline{\includegraphics[width=0.95\columnwidth]{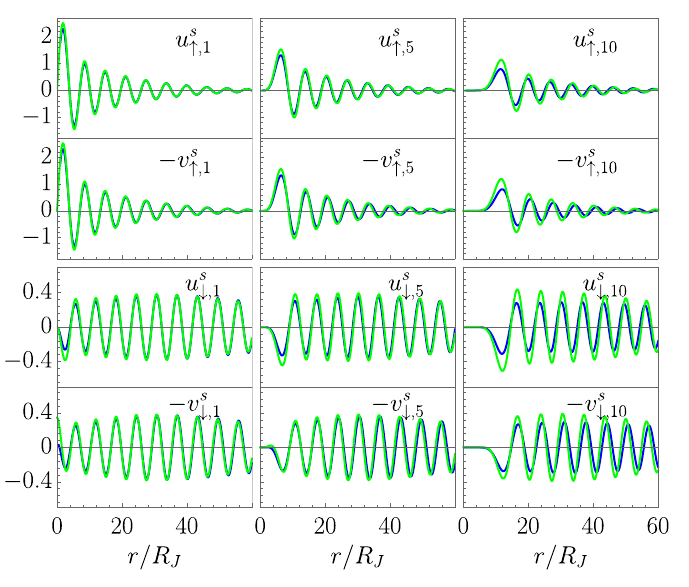}}
	\caption{Wave functions for the states localized at a vortex in the presence of a skyrmion: comparison of numerical (blue) and analytical (green) solutions for    $l=1,\,5,$ and $10$. Here we choose $R=30R_J$ and $\delta=20R_J$ (corresponding to the curve `4' in Fig. \ref{fig:genspSK}). Analytical solutions are given by Eq. \eqref{eq:MM:sol:Sk:l}. All other parameters are the same as in Fig. \ref{fig:noSk}. Wave functions are normalized according to Eq. \eqref{eq:norm:cond:m} and multiplied by  a factor $10^2R_J$.
    }
	\label{fig:lnonzero-Sk}
\end{figure}

\subsubsection{States localized near the vortex core}

To describe the second branch analytically we assume the exchange coupling~$J$ to be the largest energy scale, see condition~\eqref{eq:J_great}, and neglect the vector potential, $A_\varphi=0$, as well as the magnetization rotation, $\theta_v=0$. Then we can find the bound states $\Phi_l^v=\{u_{\uparrow,l}^v,u_{\downarrow,l}^v,v_{\downarrow,l}^v,-v_{\uparrow,l}^v\}^T$ approximately within the same WKB-type approximation, which was used for obtaining Eq.~\eqref{eq:JJ:up}, see details in Appendix~\ref{App:D}, 
\begin{align}
 u_{\uparrow,l}^v & \simeq c_{v,l} e^{-\mathcal{S}(r)}
 \Bigl [ \mathcal{J}_l\Bigl(\frac{r}{R_J}\Bigr ) +\frac{R_J\Delta}{|\alpha|}\mathcal{J}_{l-1}\Bigl(\frac{r}{R_J}\Bigr )   \Bigr ]
 \label{eq:up:WKB:lnot0}\\
 \notag
 v_{\uparrow,l}^v & 
 \simeq -c_{v,l} (\sgn \a) e^{-\mathcal{S}(r)}\Bigl [ \mathcal{J}_l\Bigl(\frac{r}{R_J}\Bigr ) -\frac{R_J\Delta}{|\alpha|}\mathcal{J}_{l+1}\Bigl(\frac{r}{R_J}\Bigr )   \Bigr ]
 ,
 \end{align}
 and
 \begin{align}
u_{\downarrow,l}^v & \simeq - 2c_{v,l}(R_JJ/\alpha) e^{-\mathcal{S}(r)} \mathcal{J}_{l+1}(r/R_J) ,
\notag
\\
v_{\downarrow,l}^v & \simeq - 2c_{v,l}(R_JJ/|\alpha|) e^{-\mathcal{S}(r)} \mathcal{J}_{l-1}(r/R_J) .
\label{eq:down:WKB:lnot0}
\end{align}
Here $\mathcal{S}(r)$ is described by Eq.~\eqref{eq:def:K} and subscript ``v'' means the bound state localized near the vortex. \color{black} It is worthwhile to mention that for $r\gg R_J \max(1,|l|)$ the particle-hole components of the wavefunctions \eqref{eq:up:WKB:lnot0}--\eqref{eq:down:WKB:lnot0} are approximately  related  as 
$v_{\ua,l}^v\simeq -\sgn\a\, u_{\ua,l}^v$ and $v_{\downarrow,l}^v\simeq -\sgn\a\, u_{\downarrow,l}^v$.
\color{black}
We compare analytic results \eqref{eq:up:WKB:lnot0} and \eqref{eq:down:WKB:lnot0} with the numerical ones in Fig. \ref{fig:lnonzero}. \color{black} Although, the result \eqref{eq:up:WKB:lnot0}-\eqref{eq:down:WKB:lnot0} is valid formally at large
distances, $r\gg R_J \max(1,|l|)$, only,
as one can see, the analytic and numerical wave functions are in good agreement for $r\gtrsim R_J$. 
\color{black}

With the explicit wave functions in hands, we can estimate the energies of the states with non-zero $l$ localized at the vortex. We choose the most straightforward way and estimate the energy as the average of the Hamiltonian (see Appendix~\ref{App:D}):
\begin{equation}
    E_l^v=\langle\Phi_l^v|H^{(l)}|\Phi_l^v\rangle \simeq - \frac{|\alpha| l}{J}    \frac{\int_0^\infty dr \Delta e^{-2\mathcal{S}(r)}\mathcal{J}_l^2(r/R_J)}{\int_0^\infty dr r e^{-2\mathcal{S}(r)}\mathcal{J}_l^2(r/R_J) }. 
    \label{eq:El:WKB}
\end{equation}
\color{black} Substituting $R_J/(\pi r)$ for $\mathcal{J}_l^2(r/R_J)$ at $r\gg |l|R_J$ and \color{black} evaluating the integrals with logarithmic accuracy, we find 
\begin{equation}
E_l^v \simeq  -l \frac{2 m \alpha^2\Delta_\infty^2}{J^2}\ln{\frac{J/(m|\alpha| \Delta_\infty)}{\max\{{\color{black} |l|}R_J,\xi_v\}}} .
\label{eq:El:main}
\end{equation}
We note that the sign of the slope $dE_l/dl$ at $l=0$ is always negative independent of the sign of the spin-orbit coupling $\alpha$. Also, we mention that the result \eqref{eq:El:main} can be understood 
on the basis of the effective superconducting gap $\Delta_{\rm eff}$, see Eq.~\eqref{eq:Delta:eff}. Indeed, except the logarithmic factor, we can write  $E_l^v\sim l \Delta_{\rm eff}^2/(J+\mu)$ which resembles a standard result for the Caroli-de Gennes-Matricon states \cite{Caroli1964}. Also, we note that the result \eqref{eq:El:main} can be understood based on Eq. \eqref{eq:El:LLL} with the spin-orbit length $\Lambda_{\rm so}$ instead of the systems size $L$ (again upto the logarithmic factor).

The green dots in Fig.~\ref{fig:sp} marks the the energies calculated with the help of the expression~\eqref{eq:El:WKB}. We emphasize good agreement between analytical expression \eqref{eq:El:WKB} and the numerically calculated energies (blue dots) even for relative large magnitudes of the angular momentum, $l\simeq \pm 15$. 

\subsection{Configuration with a skyrmion}

In the presence of skyrmion, the energy spectrum consists of two branches also. The first branch corresponds to the states localized near the edge. Provided that $L\gg R$, these states are not affected by skyrmion magnetization, $\theta_{R\delta}(r\gg R)\approx0$, and, therefore, this branch is described by Eq.~\eqref{eq:El:LLL}.

The situation is more sophisticated with the second branch because of the skyrmion angle $\theta(r)$. 
We use analytical approach for the second branch based on the rotation of the Hamiltonian, see Appendix~\ref{App:D}. 
Supposing conditions~\eqref{eq:J_great} and~\eqref{eq:Rdelta_cond}, we derive the WKB-type wave functions similar to~\eqref{eq:MM:sol:Sk} for the bound states with nonzero $l$,
\begin{equation}
\begin{aligned}
		u_{\uparrow,l}^s(r) & = \chi\Big[\underline{u}_{{\downarrow,l}}(r) \cos\frac{\theta(r)}{2} +\underline{u}_{{\uparrow,l}}(r) \sin \frac{\theta(r)}{2}
        \Big],  \\
		u_{\downarrow,l}^s(r) & = \chi\Big[\underline{u}_{{\downarrow,l}}(r) \sin\frac{\theta(r)}{2} -\underline{u}_{{\uparrow,l}}(r) \cos \frac{\theta(r)}{2}
        \Big],
        \\
		v_{\uparrow,l}^s(r) & = \chi\Big[\underline{v}_{{\downarrow,l}}(r) \cos \frac{\theta(r)}{2}
        +\underline{v}_{{\uparrow,l}}(r) \sin\frac{\theta(r)}{2} 
        \Big],
        \\
        v_{\downarrow,l}^s(r) & = \chi\Big[\underline{v}_{{\downarrow,l}}(r) \sin \frac{\theta(r)}{2}
        -\underline{v}_{{\uparrow,l}}(r) \cos\frac{\theta(r)}{2} 
        \Big] .
\end{aligned}
\label{eq:MM:sol:Sk:l}
\end{equation}
Here the subscript ``s'' means the bound state localized near the vortex and the coaxial skyrmion, and
\begin{align}
\underline{u}_{\uparrow,l}  \simeq & -(\sgn\a)\underline{v}_{\uparrow,l}  
\simeq -\frac{2 \tilde{c}_l R_J J} {\alpha} e^{{-}\mathcal{S}(r)}
\mathcal{J}_{l}\Bigl (\frac{r}{R_J}\Bigr ),
\notag \\
\underline{u}_{\downarrow,l} \simeq &  \tilde{c}_l e^{{-}\mathcal{S}(r)}   
\Bigl [ \mathcal{J}_{l+1}\Bigl (\frac{r}{R_J}\Bigr ) {+} \frac{\Delta R_J}{|\alpha|} \mathcal{J}_{l}\Bigl (\frac{r}{R_J}\Bigr ) 
\Bigr ],
\label{eq:v:dd:3} \\
\underline{v}_{\downarrow,l} \simeq &-(\sgn\a)\tilde{c}_l e^{{-}\mathcal{S}(r)}   
\Bigl [ {-}\mathcal{J}_{l-1}\Bigl (\frac{r}{R_J}\Bigr ) {+} \frac{\Delta R_J}{|\alpha|} \mathcal{J}_{l}\Bigl (\frac{r}{R_J}\Bigr ) 
\Bigr ] , \notag 
\end{align}
and $\mathcal{S}(r)$ is described by Eq.~\eqref{eq:def:K}. \color{black} We note that for $r\gg R_J\max(1,|l|)$ the particle-hole components of the wavefunctions \eqref{eq:v:dd:3} are approximately  related  as 
$v_{\ua,l}\simeq -\sgn\a\, u_{\ua,l}$ and $v_{\downarrow,l}\simeq -\sgn\a\, u_{\downarrow,l}$.
\color{black} We compare analytic results \eqref{eq:MM:sol:Sk:l}  with the numerical ones in Fig. \ref{fig:lnonzero-Sk}. As one can see, the analytic and numerical wave functions are in qualitative agreement. We note that as the magnitude of $l$ increases, the quantitative agreement becomes worse.
This may be related with the fact the results \eqref{eq:v:dd:3} are strictly valid for $r\gg R_J \max(1,|l|)$. Another possible source of the discrepancy is the increase of the absolute value of $E_l^s$ with $|l|$, while the energy is neglected in the construction of the solution \eqref{eq:v:dd:3}.

\begin{figure}[t!]
	\centerline{\includegraphics[width=0.95\columnwidth]{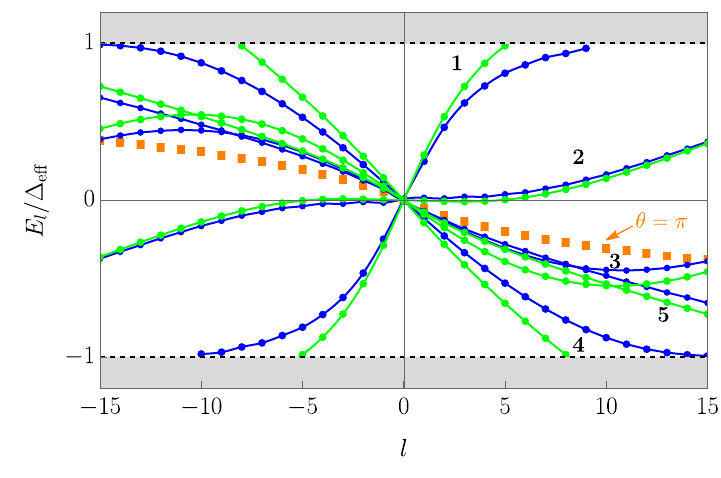}}
	\caption{The dependence of energies $E_l^s$ on $l$ for different skyrmion radii $R$: comparison of numerical (blue) and analytical (green) results, cf. Eq. \eqref{eq:El:WKB:Sk}. The numbers `1'-`5' on curves correspond to skyrmion radii $R/R_J=5, 10, 15, 30,$ and $100$, respectively. Orange points correspond to the energies $E_l^v$ for the case $\theta\equiv\pi$.  Black dashed line depicts the effective energy gap, cf. Eq. \eqref{eq:Delta:eff}. The gray bars show the regions where the states are delocalized. We use $\delta=20R_J$ and all other parameters are the same as in Fig. \ref{fig:noSk}. }
	\label{fig:genspSK}
\end{figure}

In order to estimate the energies of the states with nonzero $l$ localized at the coaxial vortex-skyrmion pair, we evaluate the average Hamiltonian. Then we find  
\begin{gather}
 E_l^s=\langle\Phi_l^s|H^{(l)}|\Phi_l^s\rangle \simeq 
\frac{l}{2m} \left [\int_0^\infty dr r e^{-2\mathcal{S}(r)}\mathcal{J}_l^2(r/R_J)\right ]^{-1}
\notag \\
{\times} 
\int_0^\infty \frac{dr}{r} e^{-2\mathcal{S}(r)}\mathcal{J}_l^2(r/R_J)
\Bigl \{ 1+\cos\theta +(\Delta \sgn\a/J)\sin \theta 
\notag \\ -2m \alpha r \Bigl[\sin \theta - (\Delta \sgn\a/J)\cos\theta\Bigl ] \Bigr \} .
 \label{eq:El:WKB:Sk}
\end{gather}
We note that for $\theta=\pi$, the result \eqref{eq:El:WKB:Sk} coincides with Eq. \eqref{eq:El:WKB}. For $R\gg \delta$, we can approximate the skyrmion angle as a step function. Then for $R\gg \delta, \Lambda_{\rm so}$ we find 
\begin{gather}
 E_l^s \simeq E_l^v+ 
l 
\frac{\int_R^\infty dr (1/r+m|\alpha| \Delta /J) e^{-2\mathcal{S}(r)}\mathcal{J}_l^2(r/R_J)}
{m \int_0^\infty dr r e^{-2\mathcal{S}(r)}\mathcal{J}_l^2(r/R_J)} .
 \label{eq:El:WKB:Sk:R}
\end{gather}
We note that in contrast to $E_l^v$ which is negative (for $\alpha>0$) the second term in Eq. \eqref{eq:El:WKB:Sk:R} is positive. Therefore, we expect that the slope $dE_l/dl$ at $l=0$ will change sign with increase of $R$ from positive to negative. 

In Fig.~\ref{fig:genspSK} we compare analytical and numerical results. We see that although there is no perfect quantitative agreement between numerics and analytics, the analytical result \eqref{eq:El:WKB:Sk} reproduces correctly dependence of $E_l$ on $l$ for various magnitudes of $R$.
As with the wave functions, the agreement between analytical and numerical results for $E_l^s$ is better for small magnitudes of $l$ and deteriorates as $|l|$ increases.

\section{Discussions and conclusions\label{Sec:Disc}}

The conditions \eqref{eq:cond} and \eqref{eq:cond:L} for the existence of the Majorana states localized at the superconducting vortex and at the edge of the system can be generalized to include the case of magnetic configurations in which the angle $\theta$ has a finite spatial derivative for $r\to \infty$. Then as it follows from the rotated Hamiltonian \eqref{eq:3:HH:R} the spin-orbit coupling appears in combination $\alpha+\partial_r\theta/(2m)$ at large $r$. Therefore, if $\lim\limits_{r\to \infty} \partial_r\theta \neq 0$ the Majorana states can exist even in the absence of the spin-orbit coupling if inequality \eqref{eq:cond} holds as was first demonstrated in Ref.~\cite{Yang2016}. 

The role of non-zero spin-orbit coupling for existence of the Majorana state localized at a superconducting vortex can be clearly seen from the analytic solutions in the absence (cf. Eqs. \eqref{eq:JJ:up}, \eqref{eq:JJ:up:pi}, \eqref{eq:JJ:up:after_rotation}) and presence (cf. Eq. \eqref{eq:MM:sol:Sk}) of a skyrmion. The spin-orbit coupling together with the superconducting gap controls the spatial localization of the Majorana state via the function $\mathcal{S}(r)$, cf. Eq. \eqref{eq:def:K}. We note that the very same function controls decay of the localized states with nonzero $l$ (cf. Eqs. \eqref{eq:up:WKB:lnot0}, \eqref{eq:down:WKB:lnot0}, \eqref{eq:El:WKB}, \eqref{eq:MM:sol:Sk:l}, and \eqref{eq:El:WKB:Sk}). We emphasize that this function $\mathcal{S}(r)$ is nothing but the function determine the decay of the standard  Caroli-de Gennes-Matricon states 
but with the effective superconducting gap \eqref{eq:Delta:eff}.

Interesting and unexpected feature of the solution \eqref{eq:MM:sol:Sk} for the Majorana state localized on a superconducting vortex in the presence of a skyrmion is that its asymptotic behavior is determined by the solution for homogeneous magnetization with the angle $\theta=\pi$ rather than $\theta=0$. Essentially, the role of a skyrmion is twofold: (i) it makes the proper boundary conditions at $r=0$ that selects solution for homogeneous magnetization $\theta=\pi$, and (ii) it rotates this solution with increase of the distance.   

We analyze the effect of the vector potential induced by both a superconducting vortex and a magnetic texture. Consistent with Ref.~\cite{Rex2019}, we find that the vector potential introduces negligible corrections to the solutions for the Majorana wave functions and can therefore be omitted. In contrast, the perturbation of the magnetic texture by the vortex-induced magnetic field can 
affect the solution for the non-skyrmion configuration.

It is instructive to compare our results with the numerical results reported in Ref.~\cite{Rex2019}. The authors used the following set of parameters: $\mu=0$, $\Delta_\infty=J/2$, $1/(m\alpha)=71 R_J$, $\Lambda_{\rm so}=123 R_J$, $R/R_J=14 \div 42$, and the exponential ansatz for the skyrmion angle, $\theta(r)=\pi \exp(-r/R)$. We note that for the above parameters, inequality \eqref{eq:Rdelta_cond:0} is satisfied, while inequality \eqref{eq:Rdelta_cond} is not.

In agreement with Ref.~\cite{Rex2019}, we find that the component of the Majorana wave function localized at the vortex-skyrmion pair, $u^s_\uparrow$, is much larger than $u^s_\downarrow$ (see Fig.~\ref{fig:Sk} and Fig.~5 of Ref.~\cite{Rex2019}). Also in agreement with Ref.~\cite{Rex2019}, the components of the wave functions of states with nonzero $l$ localized at the vortex-skyrmion pair demonstrate approximate coincidence at large distances, $u^s_{\uparrow,l} = -v^s_{\uparrow,l}$ (see Fig.~\ref{fig:lnonzero} and Fig.~5 of Ref.~\cite{Rex2019}). However, our analytical and numerical results also predict a similar relation, $u^s_{\downarrow,l} = -v^s_{\downarrow,l}$, which is not observed in the numerics of Ref.~\cite{Rex2019}.

Another point of discrepancy is the sign and magnitude of the slope $dE_l/dl$ for states localized near the rim. In our work, we find analytically and numerically that the slope is negative for both signs of $\alpha$ [cf. Eq.~\eqref{eq:El:LLL}], whereas in Ref.~\cite{Rex2019} the slope is positive. Furthermore, while in Ref.~\cite{Rex2019} the quantity $(dE_l/dl)/E_1$ coincides with our analytical estimates, the absolute value of $E_l$ is found to be 10 times smaller than predicted by our theory (see Fig.~\ref{fig:sp} and Fig.~4 of Ref.~\cite{Rex2019}). Finally, the sign of the slope $dE^s_l/dl$ for states localized at the vortex-skyrmion pair in Ref.~\cite{Rex2019} is positive, in agreement with our numerical results for the range of skyrmion radii considered there (see curves 3 and 4 in Fig.~\ref{fig:genspSK} and Fig.~4 of Ref.~\cite{Rex2019}). However, a magnitude of $E^s_l$ in Ref.~\cite{Rex2019} is approximately 10 times smaller than we find here. 

Our theory predicts that in the absence of a skyrmion, the Majorana state localized at a superconducting vortex is accompanied by a set of localized states with nonzero angular momentum $l$, separated by a gap of order $\Delta_{\rm eff}^2/J$. The presence of such states, separated from zero energy by a small energy gap [cf. Eq.~\eqref{eq:Delta:eff}], can be a detrimental factor for the manipulation of Majorana states. The presence of a skyrmion may further complicate the situation. Indeed, there exists a range of skyrmion radii for which the energies of localized states with small angular momenta become anomalously close to zero energy (see, e.g., curve 2 in Fig.~\ref{fig:genspSK}). 
Conversely, the presence of a skyrmion of an appropriate radius can increase this gap (see, e.g., curves 1 and 4 in Fig.~\ref{fig:genspSK}) and thereby facilitate the identification of different localized states. We also emphasize that the skyrmion radius is sensitive to the external magnetic field~\cite{Apostoloff2024}. In the same heterostructure, the radii of a free skyrmion and a skyrmion coupled to a vortex can differ by a factor of several times~\cite{Apostoloff2023,Xie2023}. Consequently, tuning the skyrmion size may prove to be a key element in the manipulation of Majorana states.

The localized states with nonzero angular momentum may also affect the skyrmion radius at finite temperature, since $E_l^s$, and thus the contribution from the thermal filling of these states to the free energy, depends on $R$. These states may also mediate interactions between vortex-skyrmion pairs in superconductor-chiral ferromagnet structures of the type used in recent experiments \cite{Petrovic2021,Xie2023}. 

The spectrum of states localized at a vortex-skyrmion pair can be affected by a current flow in the superconductor. Since in the considered model both parity and time-reversal symmetry are broken, a superconducting diode effect can be expected (see e.g. Ref.~\cite{Hess2023}). It would be interesting to study how the corresponding critical currents depend on the spectrum of localized states. We note that the existence of a diode effect in a superconductor-chiral ferromagnet structure was indirectly observed in Ref.~\cite{Petrovic2021}, where an asymmetry of the superconducting critical current with respect to a sign change of the magnetic field was reported.

Another interesting question that can be addressed within the WKB-type approach developed in our work is the fate of the Majorana state localized at a coaxial vortex-skyrmion pair when the skyrmion's center is shifted relative to the superconducting vortex.

To summarize, within the BdG Hamiltonian, we study the low-lying states localized at a superconducting vortex in a thin superconductor--chiral ferromagnet heterostructure with or without a skyrmion. We find a rigorous criterion for the existence of Majorana states in such a system and prove that the parity of the Majorana state localized at the vortex is opposite to the sign of spin-orbit coupling, $\eta = -\sgn\alpha$ [cf. Eqs.~\eqref{eq:cond} and \eqref{eq:cond:L}]. Under reasonable assumptions, we develop an analytical theory for the wavefunctions of the Majorana state localized at the vortex without a skyrmion [cf. Eqs.~\eqref{eq:JJ:up}, \eqref{eq:JJ:up:pi}, and \eqref{eq:JJ:up:after_rotation}], and in the presence of a coaxial skyrmion [cf. Eq.~\eqref{eq:MM:sol:Sk}], as well as for the Majorana state localized at the edge of the system. We also derive analytical expressions for the wave functions and energies of states with nonzero $l$ localized at the vortex [cf. Eqs.~\eqref{eq:up:WKB:lnot0}, \eqref{eq:down:WKB:lnot0}, and \eqref{eq:El:WKB}], and at the coaxial vortex-skyrmion pair [cf. Eqs.~\eqref{eq:MM:sol:Sk:l} and \eqref{eq:El:WKB:Sk}], as well as of states localized at the edge [cf. Eq.~\eqref{eq:El:LLL}]. All our analytical results are in good agreement with numerical solutions. Furthermore, we estimate the role of such realistic effects as the vector potential induced by the vortex, the noncollinear magnetic texture, and the perturbation of the magnetic texture due to the stray magnetic field generated by the superconducting vortex.

\begin{acknowledgements}
The authors are grateful to D. Katkov for the collaboration on the initial stage of the project. The authors acknowledge useful discussions with E. Andriyakhina, I. Gornyi, Ya. Fominov, I. Kolokolov, V. Lebedev, A. Mel'nikov, A. Samokhvalov, and M. Skvortsov. The authors would like especially to thank A. Kudlis for his kind assistance with Python coding. \color{black} The work was funded in part by the Russian Science Foundation under the grant No. 24-12-00357 (analytics for Majorana states), by Ministry of Science and Higher Education (numerical results), and by Basic research program of HSE under Grant No. HSE-BR-2025-57 (analytics for localized states). \color{black} The authors acknowledge personal support from the Foundation for the Advancement of Theoretical Physics and Mathematics``BASIS''. We acknowledge the computing time provided to us at computer facilities at L.D. Landau Institute for Theoretical Physics.
\end{acknowledgements}

\begin{center}
 \bf{DATA AVAILABILITY}  
\end{center}

The data that support the findings of this article are not
publicly available. The data are available from the authors
upon reasonable request.

\appendix

\section{Conditions for existence of the Majorana zero mode localized at the vortex core\label{App:A}}

In this section we analyze asymptotic behavior of the Majorana mode localized near the origin at $r\to 0$ and $r\to \infty$. The analyze of the limit $r\to 0$ is needed in order to formulate the correct boundary conditions.   

\subsection{The regime $r\to 0$}

In order to study the asymptotic behavior of the Majorana mode in the region near the origin, it is convenient to use the parametrization $\phi^{\eta}_{(o)}(r)=\{H_{\uparrow}(r), r H_{\downarrow}(r)\}$. Then the functions $H_{\uparrow,\downarrow}(r)$ solve the following equations  
\begin{gather}
- \partial_r^2 H_\sigma - (2-\sigma) r^{-1}\partial_r H_\sigma + X_\sigma(r) H_\sigma + Y_\sigma(r) H_{-\sigma}
\notag \\+ 2m\alpha \sigma r^{\sigma} \partial_r H_{-\sigma}=0 .
\label{eq:App:A:00}
\end{gather}
Here the functions $X_\sigma$ and $Y_\sigma$ are defined as 
\begin{align}
 X_\sigma(r) & {=} [(\sigma-1)/r] eA_\varphi(r)  {+}e^2A_\varphi^2(r) {-}2m [\mu {-}\sigma J\cos\theta(r)] ,\notag \\ 
 Y_\sigma(r) & = 2m r^{\sigma}[J \sin\theta(r)+\eta\sigma \Delta(r)-\alpha e A_\varphi(r) \notag \\
          & +(1+\sigma)\alpha r^{-\sigma}] .
\end{align}
The functions $X_\uparrow(r)$, $Y_\uparrow(r)$ have a finite limit at $r\to0$, while $X_\downarrow(r)$, $Y_\downarrow(r)$ have a logarithmic asymptotics due to $A_{\varphi}$ (see~\eqref{eq:vortex:A:r}). So, one can write at $r\to0$:
\begin{align}
    X_\downarrow(r) & =X_\downarrow^{(0)}+2eB_0\ln\frac{r}{d_S},\quad X_\uparrow(r)  =X_\uparrow^{(0)} , \notag \\
    Y_\downarrow(r) & =Y_\downarrow^{(0)}+2m \alpha eB_0\ln\frac{r}{d_S},  \quad Y_\uparrow(r) =Y_\uparrow^{(0)} ,
\end{align}
where $X_\sigma^{(0)},\,\,Y_\sigma^{(0)}$ are independent of $r$, and $B_0=\phi_0/(4\pi \lambda_L^2)\equiv H_{c1}/\ln(\lambda_L/\xi)$.

Substituting  $H_\sigma$ as a power series in $r$, $H_\uparrow=1+c_\uparrow r^2$ and $H_\downarrow=c_0+c_\downarrow r^2 \ln(r/d_S)+\tilde{c}_\downarrow r^2$, into Eqs.~\eqref{eq:App:A:00}, we find
\begin{align}
\tilde{c}_\downarrow &= \frac{X_\downarrow^{(0)} c_0+Y_\downarrow^{(0)} }{8} -m\alpha\frac{X_\uparrow^{(0)} +Y_\uparrow^{(0)}  c_0}{8}, 
\notag \\
 c_\downarrow&=\frac{eB_0}{4}\Big(c_0+m\alpha\Big),\,\,\,c_\uparrow = \frac{X_\uparrow^{(0)}  +Y_\uparrow^{(0)}  c_0}{4}.
\end{align}
Therefore, the functions $H_\sigma$ have a well-defined limit at $r=0$. The derived asymptotic expressions justify the boundary conditions~\eqref{eq:BC:origin}.

\subsection{The regime $r\to \infty$}

Now we analyze the behavior of the solution at $r\to \infty$. It is convenient to introduce  $\tilde{\phi}^{\eta}_{(o)}(r)=\sqrt{r}\phi^{\eta}_{(o)}(r)$. Then $\tilde{\phi}^{\eta}_{(o)}(r)$ obeys the following equation
\begin{gather}
\Bigl [ -\frac{1}{2m}\Bigl( \partial_r^2+\frac{1}{4r^2}-\frac{1-\sigma_z}{2r^2}\Bigr ) 
-\mu + J\sigma_z +i \eta \Delta_\infty \sigma_y \notag \\
+ i \alpha \sigma_y \partial_r + \frac{\alpha\sigma_x}{2r}
\Bigr ] \tilde{\phi}^{\eta}_{(o)} = 0 .   
\label{eq:appA:2}
\end{gather}
Let us seek its solution as an expansion in powers of $1/r$:
\begin{equation}
\tilde{\phi}^{\eta}_{(o)}(r) = e^{-Q r} \Bigl [ \begin{pmatrix}
1 \\
\zeta_Q
    \end{pmatrix} 
    + \frac{1}{r^z}\begin{pmatrix}
\beta_{Q,\uparrow}  \\
\beta_{Q,\downarrow} 
    \end{pmatrix} + \dots 
 \Bigr ]    
 \label{eq:appA:3}
\end{equation}
Here $Q$, $\zeta_Q$, $\beta_{Q,\uparrow}$, $\beta_{Q,\downarrow}$, and $z$ have to be determined. Substituting the anzats~\eqref{eq:appA:3} into Eq.~\eqref{eq:appA:2}, we find 
\begin{equation}
W_Q  \begin{pmatrix}
1 \\
\zeta_Q
    \end{pmatrix}  + \frac{1}{r^z} W_Q \begin{pmatrix}
\beta_{Q,\uparrow}  \\
\beta_{Q,\downarrow} 
    \end{pmatrix} + \frac{\alpha\sigma_x}{2r} \begin{pmatrix}
1 \\
\zeta_Q
    \end{pmatrix} +\dots = 0 ,   
    \label{eq:App:A:4}
\end{equation}
where
\begin{equation}
W_Q = \begin{pmatrix}
-\frac{Q^2}{2m}-\mu +J & \eta \Delta_\infty -\alpha Q \\
-\eta \Delta_\infty +\alpha Q & -\frac{Q^2}{2m}-\mu -J
\end{pmatrix}    
\label{eq:App:A:WQ}
\end{equation}
Equation~\eqref{eq:App:A:4} determines the exponent $z=1$. Then Eq.~\eqref{eq:App:A:4} splits into two equations
\begin{gather}
    W_Q  \begin{pmatrix}
1 \\
\zeta_Q
    \end{pmatrix}=0, \qquad W_Q \begin{pmatrix}
\beta_{Q,\uparrow}  \\
\beta_{Q,\downarrow} 
    \end{pmatrix} = - \frac{\alpha}{2}\sigma_x\begin{pmatrix}
1 \\
\zeta_Q
    \end{pmatrix} .
    \label{eq:App:A:5}
\end{gather}
In order the solution of the first equation exists, the determinant of the matrix $W_Q$ has to be zero. It 
leads to Eq.~\eqref{t:eq:Q}. We note that the left eigenvector from the kernel of $W_Q$ has the form $(1,-\zeta_Q)$, where 
\begin{equation}
   \zeta_Q = \frac{Q^2/(2m)+\mu-J}{\eta \Delta_\infty-\alpha Q} .
   \label{eq:App:A:zetaQ}
\end{equation}
As one can check, the following identity holds: $(1,-\zeta_Q)\sigma_x (1,\zeta_Q)^T=0$. It guarantees that the second equation in Eq.~\eqref{eq:App:A:5} has a solution for $\beta_{\uparrow,\downarrow}$.  

As we discussed in the main text for $\eta=-1$, Eq.~\eqref{t:eq:Q} has three roots for $Q$ with positive real part. We denote them as $Q^{(-)}_{1,2,3}$. Then we can construct asymptotic solution for the Majorana zero mode localized near the origin  ($\eta=+$) as 
\begin{gather}
    \tilde{\phi}_{+}^{(o)}(r) = \sum_{j=1}^3 c_j e^{-Q^{(-)}_j  r} \Bigl [ \begin{pmatrix}
1 \\
\zeta_{Q^{(-)}_j}
    \end{pmatrix} 
    + \frac{1}{r}\begin{pmatrix}
\beta_{Q^{(-)}_j,\uparrow}  \\
\beta_{Q^{(-)}_j,\downarrow} 
    \end{pmatrix} + \dots 
 \Bigr ]   .
\end{gather}
Therefore, we have four constants, $c_{1,2,3}$ and $c_0$, that determine the asymptotic behavior of the Majorana zero mode at $r\to \infty$ and $r\to 0$. One can fix them by matching asymptotics at some intermediate $r$ as described in the main text. 

%%%

\section{Analytic results for the wave function of the Majorana mode localized near the vortex core \label{App:B}}

In this Appendix we present details of the derivation of asymptotic behavior of the Majorana state localized near the vortex center in the presence of uniform magnetization, cf. Eqs.~\eqref{eq:JJ:up}, and discuss how non-zero $A_\varphi$ and $\theta_v$ can be taken into account. As in the main text, we assume that the energy $J+\mu$ is much larger than the superconducting order parameter and spin-orbit splitting, see Eq. \eqref{eq:J_great}.
It is convenient to introduce the following dimensionless parameters
$x=r/R_J$, $\bar{\alpha}=2m\alpha R_J$, $\bar{\Delta}(x)=\Delta(r)/(J+\mu)$, and 
$a_\varphi=eA_\varphi R_J$. Also, for a sake of simplicity, we assume that $\alpha>0$. Then Eqs.~\eqref{eq:eq:M} and~\eqref{eq:3:HH} read
\begin{align}
 \pd_x^2u_{\uparrow} & + \frac{ \pd_xu_{\uparrow}}{x}- u_{\uparrow}\Bigl [\frac{J \cos\theta(x)-\mu}{J+\mu}+a^2_\varphi(x)\Bigr ]  \notag \\
  = & {-} u_{\downarrow}
 \Bigl [\bar{\Delta}(x){-}\frac{J \sin\theta(x)}{J{+}\mu}{+}\bar{\alpha}a_\varphi(x)\Bigr ]{+}\bar{\alpha}\Bigl (\pd_xu_{\downarrow}{+}\frac{u_{\downarrow}}{x}\Bigr )  ,\notag
 \\
  \pd_x^2u_{\downarrow} & + \frac{ \pd_xu_{\downarrow}}{x}+ u_{\downarrow}\Bigl [\frac{J\cos\theta(x)+\mu}{J+\mu}-\Bigl (\frac{1}{x}-a_\varphi(x)\Bigr)^2
  \Bigr ]  \notag \\
   = & u_{\uparrow} \Bigl [\bar{\Delta}(x)+\frac{J \sin\theta(x)}{J+\mu}-\bar{\alpha}a_\varphi(x)\Bigr ] -\bar{\alpha} \pd_xu_{\uparrow} .
\label{eq:uu:f}
\end{align}

\subsection{The case of the uniform magnetization, $\theta=0$, and large Pearl length, $A_\varphi=0$.}

In this section we construct the solution for the wave function of the Majorana state localized near the center of the vortex for the uniform magnetization. We start from the case of infinite Pearl length, $\lambda\to \infty$. Then we can set $\theta$ and $a_\varphi$ to zero in Eqs.~\eqref{eq:uu:f}. A key idea is that since for $\alpha=\Delta_\infty=0$ characteristic equation \eqref{t:eq:Q} has solution $Q=\pm i/R_J$ corresponding to a fast (since $R_J$ is the shortest length scale) oscillations of the wave function, we are seeking solution for $u^v_{\uparrow,\downarrow}$ as
\begin{gather}
u^v_{\uparrow,\downarrow}(x) =\frac{A^v_{\uparrow,\downarrow}(x)}{\sqrt{x}} e^{i x} + {\rm c.c},
\label{eq:uv:as:Av:app}
\end{gather}
where $A^v_{\uparrow,\downarrow}(x)$ are slow functions of $x$. Then we obtain the following (yet exact) equations for $A^v_{\uparrow,\downarrow}(x)$: 
\begin{align}
 \partial^2_xA^v_{\uparrow}&+2 i \partial_x A^v_{\uparrow} + \frac{A^v_{\uparrow}}{4x^2} - \frac{2J}{J+\mu} A^v_{\uparrow}  \notag \\
 & = - \bar{\Delta}(x)  A^v_{\downarrow}
 +\bar{\alpha} \left (\partial_x A^v_{\downarrow}+i A^v_{\downarrow} +\frac{A^v_{\downarrow}}{2x}\right ) , \\
 \partial^2_x A^v_{\downarrow}&+2 i \partial_x A^v_{\downarrow} - \frac{3A^v_{\downarrow}}{4x^2} \notag \\
  & = \bar{\Delta}(x) A^v_{\uparrow}
 -\bar{\alpha} \left (\partial_x A^v_{\uparrow}+i A^v_{\uparrow} -\frac{A^v_{\uparrow}}{2x}\right ) .
\end{align}
Our aim is to find the solutions of the above equations at $x\gg 1$. The slowness of the functions $A^v_{\uparrow,\downarrow}$ means that $|\partial_x A^v_{\uparrow,\downarrow}|\ll |A^v_{\uparrow,\downarrow}|$. This allows us to derive simplified first-order differential equations:
\begin{equation}
\begin{split}
 A^v_{\uparrow}(x)  & \simeq \frac{J+\mu}{2J} \left [\bar{\Delta}(x)
 -i \bar{\alpha}-\frac{\bar{\alpha}}{2x}\right ] A^v_{\downarrow} ,\\
 2 i \partial_x A^v_{\downarrow}  & \simeq \left [ \bar{\Delta}(x)
 -i \bar{\alpha}+\frac{\bar{\alpha}}{2x} \right ] A^v_{\uparrow}  +\frac{3}{4x^2}A^v_{\downarrow} .
 \end{split}
\label{eq:AA:ud}
\end{equation}
Solving Eqs.~\eqref{eq:AA:ud}, we find for $x\gg 1$:
\begin{align}
A^v_\uparrow & \simeq b_0 e^{-\mathcal{S}(r)} e^{i \mathcal{W}(r)}
\Biggl [
 \bar{\Delta}(x) \Bigl (1  + \frac{3 i}{8 x}\Bigr )
 -i \bar{\alpha} \Bigl ( 1 - \frac{i}{8 x} 
 \Bigr ) \Biggr ] ,
\notag \\
A^v_\downarrow & \simeq  \frac{2J b_0}{(J+\mu)} e^{-\mathcal{S}(r)} e^{i \mathcal{W}(r)}
\Bigl [ 1 
+\frac{3i}{8x} \Bigr ] .  
\label{eq:A:down:f}
\end{align}
where $\mathcal{S}(r)$ is given by Eq. \eqref{eq:def:K} and 
\begin{equation}
\mathcal{W}(r)=\frac{m \alpha^2 r}{2J R_J} - \frac{1}{4J(J+\mu) R_J}\int\limits_0^r dr_1 \Delta^2(r_1) .
\label{eq:W(r)}
\end{equation}
The function $\mathcal{W}(r)$ describes small correction (of the order of $\max(\bar{\Delta}^2,\bar{\alpha}^2$)) to the period of fast oscillations. We neglect this correction as it was done in the main text of the paper. 

The asymptotic expressions \eqref{eq:A:down:f} determine the asymptotic expressions for $u^v_{\uparrow,\downarrow}$. The latter can be compared with the asymptotes of the Bessel functions of the first kind at \color{black} $x\gg \max(1,|\nu|)$ \color{black} \cite{Bateman1953}:
\begin{equation}
\mathcal{J}_\nu(x)\simeq \sqrt{\frac{2}{\pi x}}\re e^{i (x-\pi (2\nu+1)/4)} 
\Bigl [ 1+i \frac{4\nu^2-1}{8x}\Bigr ] .    
\label{eq:Jn:a}
\end{equation}
As one can check the choice $b_0=e^{i \pi /4} \tilde{c}_v \sqrt{1/2\pi}/\bar{\alpha}$ allows us to express the asymptotic expressions for $u^v_{\uparrow,\downarrow}$ in terms of the Bessel functions as it is given by Eq. \eqref{eq:JJ:up}.

\subsection{The case of the uniform magnetization, $\theta=0$, and a finite Pearl length, $A_\varphi\neq 0$.}

In this section we demontrate how the presence of non-zero vector potential affects the solution for the wave function of the Majorana state localized at the vortex. Using the parametrization \eqref{eq:uv:as:Av:app}, we find the following (yet exact) equations: 
\begin{align}
 \partial_x^2 A^v_{\uparrow}&+2 i \partial_x A^v_{\uparrow} + \frac{A^v_{\uparrow}}{4x^2} - a^2_\varphi(x) A^v_{\uparrow}- \frac{2J}{J+\mu} A^v_{\uparrow}  \notag \\ = & - [\bar{\Delta}(x) +\bar{\alpha}a_\varphi(x) ] A^v_{\downarrow}
 +\bar{\alpha} \left (\partial_x A^v_{\downarrow}+i A^v_{\downarrow} +\frac{A^v_{\downarrow}}{2x}\right ) , \notag \\
 \partial_x^2 A^v_{\downarrow}&+2 i \partial_x A^v_{\downarrow} + \frac{A^v_{\downarrow}}{4x^2}-\left (\frac{1}{x}-a_\varphi(x)\right )^2A^v_{\downarrow} \notag \\
 = & [\bar{\Delta}(x) {-}\bar{\alpha}a_\varphi(x) ] A^v_{\uparrow}
 {-}\bar{\alpha} \left (\partial_x A^v_{\uparrow}{+}i A^v_{\uparrow}{-}\frac{A^v_{\uparrow}}{2x}\right ) .
\end{align}
Repeating the analysis described in the previous section, we obtain
\begin{gather}
 A^v_{\uparrow} \simeq 
 b_0  e^{-\mathcal{S}(r)} e^{i \mathcal{W}_{A_\varphi}(r)}
 \Biggl \{ 
 \bar{\Delta}(x)\Bigl [1  + \frac{3 i}{8 x}
 \Bigr ] \notag \\
 -i \bar{\alpha} \Bigl [ 1 - \frac{i}{8 x}
 +i a_\varphi(x) 
 \Bigr ] \Biggr \}  ,
 \label{eq:A:up:f}
\end{gather}
and 
\begin{equation}
A^v_{\downarrow} \simeq 
 \frac{2J b_0}{J+\mu}  e^{-\mathcal{S}(r)} e^{i \mathcal{W}_{A_\varphi}(r)}  
 \Bigl [1  + \frac{3 i}{8 x}
 \Bigr ] .
\end{equation}
Here we introduce 
\begin{equation}
\mathcal{W}_{A_\varphi}(r)= \mathcal{W}(r) - \frac{e R_J}{2}\int\limits_r^\infty d r_1 A_\varphi(r_1) \Bigl [e A_\varphi(r_1) +\frac{2}{r_1}\Bigr ]
.   
\end{equation}
In order to estimate contibutions due to the vector potential, we consider its part, generated by a Pearl vortex, $A_\varphi^{v}$. As we have discussed in the main text, the part of the vector potential due to magnetic texture, $A_\varphi^m$, is typically smaller than $A_\varphi^{v}$. Then we find that the difference $\mathcal{W}_{A_\varphi}(r)-\mathcal{W}(r)$ can be estimated to be of the order of $(R_J/\lambda)\ln(1+\lambda/r)$. Therefore, the vector potential induces small (of the order of $R_J/\lambda$), almost constant phase shift to the fast oscillations. More serious effect is related with the term $a_\varphi$ in the last line of Eq. \eqref{eq:A:up:f}. Although at $r\ll\lambda$, $a_\varphi\ll 1/x$, for $r\gg \lambda$, we have $a_\varphi=-1/(2x)$ such that it affects the long distance asymptotic behavior of $A^v_{\uparrow}$. In particular, for $r\gg \lambda$, $u_\uparrow^v$ cannot be described by Eq. \eqref{eq:JJ:up}.

\subsection{The case of the presence of a skyrmion}

Now we describe how to construct the wave function of the Majorana state localized at the vortex in the presence of the skyrmion. We consider the rotated Hamiltonian~\eqref{eq:3:HH:R}. We assume that the Pearl length is large enough and, thus, omit the vector potential. 
As before, we  seek the zero eigenmodes of the rotated Hamiltonian~\eqref{eq:3:HH:R} at $x\gg 1$ in the following form:
\begin{equation}
\underline{u}_{{\uparrow,\downarrow}}=\underline{A}_{\uparrow,\downarrow}(x) e^{i x}/\sqrt{x}+{\rm c.c}. 
\end{equation}
Then the functions $\underline{A}_{\uparrow,\downarrow}(x)$ satisfy 
\begin{widetext}
\begin{align}
 \partial_x^2\underline{A}_{\uparrow}&+2 i \partial_x\underline{A}_{\uparrow} + \frac{\underline{A}_{\uparrow}}{4x^2} - \frac{1-\cos\vartheta}{2x^2} \underline{A}_{\uparrow} 
 - \frac{1}{4}(2\bar{\alpha}+\partial_x {\vartheta})\partial_x {\vartheta} \underline{A}_{\uparrow}
 - \bar{\alpha} \frac{\sin \vartheta}{2x} \underline{A}_{\uparrow} 
\notag
 \\
& =   \frac{\sin \vartheta}{2x^2} \underline{A}_{\downarrow} - \Bigl (\bar{\Delta}(x)-\frac{\partial_x^2{\vartheta}}{2} \Bigr ) \underline{A}_{\downarrow}
 +\bar{\alpha} \left (\partial_x\underline{A}_{\downarrow}+i \underline{A}_{\downarrow} +\frac{\cos\vartheta}{2x}\underline{A}_{\downarrow} \right )
 +\partial_x{\vartheta} \left (\partial_x\underline{A}_{\downarrow}+i \underline{A}_{\downarrow} \right ), 
\end{align}
 \begin{align}
\partial_x^2\underline{A}_{\downarrow}&+2 i \partial_x \underline{A}_{\downarrow} + \frac{\underline{A}_{\downarrow}}{4x^2}-\frac{1+\cos\vartheta}{2x^2} \underline{A}_{\downarrow} 
   - \frac{2J}{J+\mu} \underline{A}_{\downarrow} - \frac{1}{4}(2\bar{\alpha}+\partial_x{\vartheta})\partial_x{\vartheta} \underline{A}_{\downarrow} + \bar{\alpha} \frac{\sin \vartheta}{2x}
 \underline{A}_{\downarrow}
 \notag \\
 & = \frac{\sin \vartheta}{2x^2}  \underline{A}_{\uparrow}+
 \left (\bar{\Delta}(x) -\frac{\partial_x^2{\vartheta}}{2} \right ) \underline{A}_{\uparrow}
 -\bar{\alpha} \left (\partial_x\underline{A}_{\uparrow}+i \underline{A}_{\uparrow} -\frac{\cos\vartheta}{2x}\underline{A}_{\uparrow}\right )
 -\partial_x{\vartheta} \left (\partial_x\underline{A}_{\uparrow}+i \underline{A}_{\uparrow} \right ) .
\end{align}
\end{widetext}
Next, within the same assumptions as before, we obtain for the `down' component
\begin{gather}
\underline{A}_{\downarrow} \simeq -\frac{J+\mu}{2J} \Bigl [\bar{\Delta}_\vartheta-i\bar{\alpha}_\vartheta+\bar{\alpha}\frac{\cos\vartheta}{2x} +\frac{\sin\vartheta}{2x^2}\Bigr ]\underline{A}_{\uparrow} .   
\label{app:eq:Auu:110}
\end{gather}
where $\bar{\Delta}_\vartheta=\bar{\Delta}-\partial_x^2\vartheta/2$ and $\bar{\alpha}_\vartheta=\bar{\alpha}+\partial_x\vartheta$.
The `up' component satisfies the following equation:
\begin{align}
 \partial_x\underline{A}_{\uparrow} \simeq & \frac{i}{2} \Bigl [\bar{\Delta}_\vartheta-i\bar{\alpha}_\vartheta-\bar{\alpha}\frac{\cos \vartheta}{2x}-\frac{\sin \vartheta}{2x^2}\Bigr ]\underline{A}_{\downarrow}  -\frac{i}{2}\Bigl [\frac{1{-}2\cos\vartheta}{4x^2}
 \notag \\
 &+\frac{\bar{\alpha}\sin\vartheta}{2x}+\frac{1}{4}(2\bar{\alpha}+\partial_x {\vartheta})\partial_x {\vartheta}\Bigr] \underline{A}_{\uparrow}  .
 \label{app:eq:Add:110}
\end{align}

Solving Eqs. \eqref{app:eq:Auu:110} and \eqref{app:eq:Add:110}, we find
\begin{align}
\underline{A}_{\uparrow} & \simeq \frac{2J b_0}{J+\mu} e^{-\mathcal{S}_\vartheta(r)}e^{i\mathcal{W}_\vartheta(r)} 
   \left ( 1 + i \frac{3\sin^2\frac{\vartheta}{2}-\cos^2\frac{\vartheta}{2}}{8x}\right ) , \notag \\
   \underline{A}_{\downarrow} & \simeq - b_0 e^{-\mathcal{S}_\vartheta(r)}e^{i\mathcal{W}_\vartheta(r)} 
   \Biggl [ \bar{\Delta}_{\vartheta} \left (1 + i \frac{3\sin^2\frac{\vartheta}{2}-\cos^2\frac{\vartheta}{2}}{8x}\right )\notag \\
   &
- i \bar{\alpha}_{\vartheta} \left (1 + i \frac{3\cos^2\frac{\vartheta}{2}-\sin^2\frac{\vartheta}{2}
-\frac{4\pd_x\vartheta \cos\vartheta}{\bar{\a}_{\vartheta}}
}{8x}\right )\Biggr ] .
\label{app:eq:Auu:112} 
\end{align}
Here we introduce the following functions
\begin{equation}
 \mathcal{S}_\vartheta(r)=\frac{m}{J}\int\limits_0^r dr_1 \left (\alpha+\frac{\partial_r\vartheta}{2m}\right )\left (\Delta(r_1)-\frac{\partial_r^2\vartheta}{4m}\right )  ,
 \label{eq:Svartheta}
\end{equation}
and 
\begin{align}
 \mathcal{W}_\vartheta(r) &= \frac{m}{2J R_J}
 \int\limits_0^r dr_1 \left (\alpha+\frac{\partial_r\vartheta}{2m}\right )^2
 \notag \\
& - \frac{1}{4J(J+\mu) R_J}\int\limits_0^r dr_1 \Delta(r_1) \left (\Delta(r_1)-\frac{\partial_r^2\vartheta}{2m}\right ) \notag \\
& - \frac{1}{4(J+\mu)R_J}\int_0^rdr_1\left(\a+\frac{\pd_r\vartheta}{4m}\right)\pd_r\vartheta\notag\\
& - \frac{m\alpha R_J}{2} \int\limits_0^r dr_1 \frac{\sin \vartheta(r_1)}{r_1} . 
\label{app:eq:Auu:113} 
\end{align}
We note that the presence of the skyrmion results in a modification of the phase of the solution. Although in the main text we neglect this effect, nevertheless, this change of the phase is visible in numerical solution, see Fig. \ref{fig:Sk}. Below we assume $\pd_x\vartheta \ll \bar{\a},\bar{\D}$.

The asymptotic expressions \eqref{app:eq:Auu:112} determine the asymptotic expressions for $\underline{u}_{\uparrow,\downarrow}$. The latter can be compared with the asymptotes  at \color{black} $x\gg \max(1,|\nu|)$ \color{black} for the Bessel functions of the first kind, cf. Eq. \eqref{eq:Jn:a}, and for the Bessel functions of the second kind \cite{Bateman1953}:
\begin{equation}
\mathcal{Y}_\nu(x)\simeq \sqrt{\frac{2}{\pi x}}\im e^{i (x-\pi (2\nu+1)/4)} 
\Bigl [ 1+i \frac{4\nu^2-1}{8x}\Bigr ] .     
\label{eq:Yn:a}
\end{equation}
As one can check, the choice $b_0=-e^{-i \pi /4} \tilde{c}_v/(\bar{\alpha} \sqrt{\pi})$ allows us to express the asymptotic expressions for $\underline{u}_{\uparrow,\downarrow}$ in terms of the Bessel functions:
\begin{equation}
\underline{u}_\uparrow {\simeq} {-}\frac{2 \tilde{c}_v R_J J} {\alpha} e^{{-}\mathcal{S}(r)}\Bigl [
\mathcal{J}_0(r/R_J) \cos^2\frac{\vartheta}{2} {-} \mathcal{Y}_1(r/R_J) \sin^2\frac{\vartheta}{2} \Bigr ] 
\end{equation}
and
\begin{align}
 \underline{u}_\downarrow  & \simeq \tilde{c}_v  e^{-\mathcal{S}(r)}  
\Biggl \{
\mathcal{J}_1(r/R_J) \cos^2\frac{\vartheta}{2} +\mathcal{Y}_0(r/R_J) \sin^2\frac{\vartheta}{2}
\notag \\
 {+} & \frac{R_J \Delta} {\alpha} \Bigl [
\mathcal{J}_0(r/R_J) \cos^2\frac{\vartheta}{2} {-} \mathcal{Y}_1(r/R_J) \sin^2\frac{\vartheta}{2} \Bigr ]   
\Biggr \} .
\end{align}
 We note that at $x\gtrsim 10$, the functions $\mathcal{Y}_{0,1}(x)$ deviate from the functions $\pm \mathcal{J}_{1,0}(x)$ by less than $10^{-2}$, respectively. So, for $x\gtrsim 10$, we use the simplified expressions $\underline{u}_\uparrow\simeq u_\uparrow^{v,\pi}$ and $\underline{u}_\downarrow\simeq u_\downarrow^{v,\pi}$ 
that lead to Eqs. \eqref{eq:MM:sol:Sk:l} and \eqref{eq:v:dd:3}.

\section{Asymptotic behavior of the Majorana zero mode localized near the edge \label{App:C}}

In this Appendix we study the behavior of the Majorana zero mode localized near the edge. We assume that the system size is large enough in order to set $A_\varphi(r)$ and $\theta(r)$ to zero in the Hamiltonian~\eqref{eq:3:HH}. Then $\tilde{\phi}_{+1}^{(o)}(r)$ obeys Eq.~\eqref{eq:appA:2}. As discussed above in Appendix~\ref{App:A}, we can construct the solution as 
\begin{equation}
\tilde{\phi}_{+1}^{(o)}(r) = \sum_{j=1}^3 c_j e^{-Q^{(+)}_j  r}  \begin{pmatrix}
1 \\
\zeta_{Q^{(+)}_j}
    \end{pmatrix}
    \label{eq:App:A:pp:an}
\end{equation}
where $Q^{(+)}_j=-Q^{-}_j$ and, consequently, $\re Q^{(+)}_j<0$. In order to satisfy the boundary conditions~\eqref{eq:BC:Infty}, the yet unknown parameters $c_{1,2,3}$ and $c_\infty$ have to satisfy the following linear equation:
\begin{gather}
 M_\zeta M_s  
 \begin{pmatrix}     c_1 \\     c_2 \\     c_3 \\     c_\infty \end{pmatrix} 
= 
\begin{pmatrix}
  0 \\
  0\\
-1 \\
0
\end{pmatrix}  ,
\label{eq:App:A:solS}
\end{gather}
where the matrices $M_\zeta$ and $M_s$ read
\begin{align}
M_\zeta  &= \begin{pmatrix}
1 & 1 & 1 & 0 \\
\zeta_{Q^{(+)}_1} & \zeta_{Q^{(+)}_2}  & \zeta_{Q^{(+)}_3}  & 0 \\
Q^{(+)}_1 & Q^{(+)}_2  & Q^{(+)}_3  & 0 \\
Q^{(+)}_1 \zeta_{Q^{(+)}_1}  & Q^{(+)}_2 \zeta_{Q^{(+)}_2}  & Q^{(+)}_3 \zeta_{Q^{(+)}_3}  & 1
\end{pmatrix} , \notag
\\
M_s  &= 
\begin{pmatrix}
 e^{-Q^{(+)}_1 L} & 0 & 0 & 0 \\
 0 & e^{-Q^{(+)}_2 L} & 0 & 0\\
 0 & 0 & e^{-Q^{(+)}_3 L} & 0 \\
 0 & 0 & 0 & 1 
\end{pmatrix} .
\end{align} 
The determinant of the product $M_\zeta M_s$ is given as
\begin{gather}
    \det (M_\zeta M_s)  =  e^{-(Q^{({+})}_1+Q^{(+)}_2+Q^{(+)}_3) L} 
 \frac{(Q^{(+)}_3-Q^{(+)}_2)}{(\Delta-\alpha Q^{(+)}_1)}\notag \\
 {\,}\hspace{.5cm}\frac{(Q^{(+)}_1{-}Q^{(+)}_3)}{(\Delta{-}\alpha Q^{(+)}_2)}
 \frac{(Q^{(+)}_2{-}Q^{(+)}_1)}{(\Delta{-}\alpha Q^{(+)}_3)}
\Bigl (\alpha^2 (\mu{-}J){+}\frac{\Delta^2_\infty}{2m}\Bigr ) .
\end{gather}
Next, Eq.~\eqref{eq:App:A:solS} can be solved explicitly, such that we find
\begin{align}
c_\infty  &{=}\frac{\alpha[
Q^{(+)}_1Q^{(+)}_2Q^{(+)}_3{+}2m(J{-}\mu) (Q^{(+)}_1{+}Q^{(+)}_2{+}Q^{(+)}_3)
]}{2m[2m \alpha^2 (\mu-J)+\Delta^2_\infty]}\notag \\
-& \frac{\Delta_\infty[2m(J{-}\mu){+}Q^{(+)}_1Q^{(+)}_2{+}Q^{(+)}_2Q^{(+)}_3{+}Q^{(+)}_3Q^{(+)}_1]}{2m[2m \alpha^2 (\mu-J)+\Delta^2_\infty]} , \notag
\end{align}
\begin{align}
c_1 & = 
\frac{\Delta_{\infty}(Q^{(+)}_2+Q^{(+)}_3)-\alpha [2m(J-\mu)+ Q^{(+)}_2Q^{(+)}_3]}{(Q^{(+)}_1-Q^{(+)}_2)(Q^{(+)}_1-Q^{(+)}_3)}\notag \\
& \times e^{Q_1^{(+)}L} \frac{(\Delta_\infty-\alpha Q^{(+)}_1)}{2m \alpha^2 (\mu-J)+\Delta^2_\infty} ,\notag 
\end{align}
\begin{align}
c_2 & = 
\frac{\Delta_{\infty}(Q^{(+)}_1+Q^{(+)}_3)-\alpha [2m(J-\mu)+ Q^{(+)}_1Q^{(+)}_3]}{(Q^{(+)}_2-Q^{(+)}_1)(Q^{(+)}_2-Q^{(+)}_3)}\notag \\
& \times e^{Q_2^{(+)}L} \frac{(\Delta_\infty-\alpha Q^{(+)}_2)}{2m \alpha^2 (\mu-J)+\Delta^2_\infty} , \notag 
\end{align}
\begin{align}
c_3 & = 
\frac{\Delta_{\infty}(Q^{(+)}_1+Q^{(+)}_2)-\alpha [2m(J-\mu)+ Q^{(+)}_1Q^{(+)}_2]}{(Q^{(+)}_3-Q^{(+)}_1)(Q^{(+)}_3-Q^{(+)}_2)}\notag 
\\
& \times e^{Q_3^{(+)}L} \frac{(\Delta_\infty-\alpha Q^{(+)}_3)}{2m \alpha^2 (\mu-J)+\Delta^2_\infty} .
\label{eq:App:A:sol:cc}
\end{align}
Thus, 
Eqs.~\eqref{eq:App:A:pp:an} and~\eqref{eq:App:A:sol:cc} determine fully the structure of the solution of Majorana state localized near the edge of a disk. 

Let us introduce the following quantities, see Eq. 
\begin{equation}
I_{ab}=- \sum_{j,j'=1}^3 \frac{c_j c_{j'}}{Q^{(+)}_j+Q^{(+)}_{j'}} \zeta^a_{Q^{(+)}_j}\zeta^b_{Q^{(+)}_{j'}} e^{-Q^{(+)}_jL-Q^{(+)}_{j'}L}   
\label{eq:III:def}
\end{equation}
For $J-\mu\gg m\alpha^2, \Delta_\infty$, we can use~\eqref{eq:El:LLL}, 
\begin{equation}
\begin{split}
Q^{(+)}_1 & \simeq -\sqrt{2m (J-\mu)} - m \alpha \Delta_\infty/J ,\\
Q^{(+)}_{2,3} & \simeq \pm i \sqrt{2m (J+\mu)} - m \alpha \Delta_\infty/J .
\end{split}
\end{equation}
Then we find 
\begin{align}
I_{00} & \simeq \frac{1}{2\bar{\Delta}_\infty\bar{\alpha}}\left (\frac{J+\mu}{J}\right)^{3/2}
\left[\left (\frac{J+\mu}{J}\right)^{1/2}\bar{\Delta}_\infty^2+\bar{\alpha}^2 \right ]
\notag \\
& \times \left [ \left (\frac{J+\mu}{J}\right)^{1/2}\bar{\Delta}_\infty^2+2\bar{\Delta}_\infty \bar{\alpha}+ \bar{\alpha}^2\frac{J-\mu}{[J(J+\mu)]^{1/2}} \right ] ,
\end{align} 
\begin{gather}
   I_{11}  \simeq \frac{2}{\bar{\alpha}\bar{\Delta}_\infty} 
   \left (\frac{J}{J+\mu}\right)^{1/2}
   \left[ \bar{\Delta}_\infty+ \bar{\alpha} \left (\frac{J-\mu}{J+\mu}\right )^{1/2}\right ]
\end{gather}
and $I_{01}\simeq - \Delta_{\infty} I_{11}/(2J)$. Using this expressions we obtain Eq.~\eqref{eq:El:LLL} in the main text.

\section{Analytical estimates for the eigen functions and eigen energies of states with non-zero $l$ localized at the vortex \label{App:D}}

In this Appendix we describe details of constructing solution for the wave functions of the states with non-zero angular momentum $l$ localized at the superconducting vortex. 
We employ the same WKB-type approach as we use in Appendix~\ref{App:B} for constructing the wave functions of the Majorana state.

\subsection{The case of the uniform magnetization, $\theta=0$}

We start from rewriting Eq. \eqref{eq:5.0} explicitly for the wave function components in terms of dimensionless parameters:
\begin{align}
			\partial_x^2u_{\uparrow}  +\frac{1}{x}\partial_x u_\uparrow& -\frac{l^2}{x^2}u_\uparrow-u_\uparrow\frac{J-\mu}{J+\mu}+\bar{E}_lu_\uparrow 
            \notag \\
            &=\bar{\a}(\partial_x u_\downarrow+\frac{l+1}{x}u_\downarrow)+\bar{\D}(x)v_\downarrow ,
            \label{eq:D1}\\
			\partial_x^2u_{\downarrow}  +\frac{1}{x} \partial_x u_\downarrow & -\frac{(l+1)^2}{x^2}u_\downarrow+(1+\bar{E}_l)u_\downarrow\notag \\
            &=-\bar{\a}(\partial_x u_\uparrow-\frac{l}{x}u_\uparrow)-\bar{\D}(x)v_\uparrow  ,\label{eq:D2}\\
			\partial_x^2 v_{\downarrow} +\frac{1}{x}\partial_x v_\downarrow & -\frac{(l-1)^2}{x^2}v_\downarrow+(1-\bar{E}_l)v_\downarrow\notag \\
            &=-\bar{\a}(\partial_x v_\uparrow+\frac{l}{x}v_\uparrow)-\bar{\D}(x)u_\uparrow , \label{eq:D3}\\
			\partial_x^2 v_{\uparrow}  +\frac{1}{x}\partial_x v_\uparrow & -\frac{l^2}{x^2}v_\uparrow-v_\uparrow\frac{J-\mu}{J+\mu}-\bar{E}_lv_\uparrow\notag \\ &=\bar{\a}(\partial_x v_\downarrow+\frac{1-l}{x}v_\downarrow)+\bar{\D}(x)u_\downarrow ,
             \label{eq:D4}
		\end{align}
where $\bar{E}_l=E_l/(J+\mu)$. For the reasons explained before we omit $A_\varphi$ and $\theta_v$. Next, at \color{black} $x\to \infty$ \color{black} we seek solutions of Eqs. \eqref{eq:D1} - \eqref{eq:D4} in the following form 
\begin{equation}
 u_{\uparrow,\downarrow}=\frac{A_{\uparrow,\downarrow}}{\sqrt{x}}e^{ix}+{\rm c.c}, \quad
 v_{\uparrow,\downarrow}=\frac{B_{\uparrow,\downarrow}}{\sqrt{x}}e^{ix}+{\rm c.c} .
\label{eq:u:v:asymp}
\end{equation}
Here $A_{\uparrow,\downarrow}$ and $B_{\uparrow,\downarrow}$ are slow functions of $x$. Neglecting derivatives of slow functions,  from Eqs. \eqref{eq:D1} - \eqref{eq:D4} we find the following relations:
\begin{equation}
 A_{\uparrow}  = - \frac{J+\mu}{2J} \Bigl [ \bar{\alpha} \Bigl (i A_{\downarrow} + \frac{2l+1}{2x} A_{\downarrow} \Bigr ) +\bar{\Delta} B_{\downarrow} \Bigr ] ,\label{eq:DD:1} 
 \end{equation}
 \begin{equation}
 B_{\uparrow}  = - \frac{J+\mu}{2J} \Bigl [ \bar{\alpha} \Bigl (i B_{\downarrow} + \frac{1-2l}{2x} B_{\downarrow} \Bigr ) +\bar{\Delta} A_{\downarrow} \Bigr ] ,
 \label{eq:DD:2} 
\end{equation}
and the following first order differential equations:
\begin{align}
2i\partial_x A_{\downarrow} & 
{-}\frac{4(l{+}1)^2{-}1}{4x^2} 
A_{\downarrow}
{=} \bar{\alpha} \Bigl [
\frac{2l{+}1}{2x}{-}i \Bigr ] A_{\uparrow}{-}\bar{\Delta} B_{\uparrow} ,
\label{eq:DD:3} \\
2i\partial_x B_{\downarrow} & 
{-}\frac{4(l{-}1)^2{-}1}{4x^2} 
B_{\downarrow}
{=}
\bar{\alpha} \Bigl [
\frac{1{-}2l}{2x}{-}i \Bigr ] B_{\uparrow} {-}\bar{\Delta} A_{\uparrow} .
\label{eq:DD:4} 
\end{align}
Here we neglect the energy $\bar{E}_l$ since as we will demonstrate below, $|\bar{E}_l|\ll \max\{\bar{\alpha}^2,\bar{\Delta}_\infty^2\}$.

Next, substituting Eqs. \eqref{eq:DD:1} and \eqref{eq:DD:2}, into Eqs. \eqref{eq:DD:3} and \eqref{eq:DD:4}, we obtain 
\begin{widetext}
\begin{equation}
\begin{pmatrix}
 \partial_x A_{\downarrow} \\
 \partial_x B_{\downarrow}
\end{pmatrix} 
= -\frac{i}{2} \begin{pmatrix}
\frac{J+\mu}{2J}(\bar{\Delta}^2-\bar{\alpha}^2)  
+ \frac{4(l+1)^2-1}{4x^2} & 2i \bar{\alpha}\bar{\Delta} \frac{J+\mu}{2J}\\
 2i \bar{\alpha}\bar{\Delta} \frac{J+\mu}{2J} & 
 \frac{J+\mu}{2J}(\bar{\Delta}^2-\bar{\alpha}^2)  
 + \frac{4(l-1)^2-1}{4x^2} 
    \end{pmatrix}
    \begin{pmatrix}
 A_{\downarrow} \\
 B_{\downarrow}
\end{pmatrix} 
\label{eq:D66}
\end{equation}
\end{widetext}
\color{black} 
Next, we 
 neglect the terms $\sim \left(\bar{\Delta}^2-\bar{\alpha}^2\right)$, which lead to a small phase correction (see Appendix~\ref{App:B}), we obtain the following solutions:
 \color{black}
\begin{align}
A_{\downarrow} & \simeq  -\tilde{c}_0 \frac{2J}{J+\mu} e^{-\mathcal{S}(x)}\Bigl [ 1+ i \frac{4(l+1)^2-1}{8x}\Bigr ]    \label{eq:DD:11}\\
B_{\downarrow} & \simeq (\sgn \a)\tilde{c}_0 \frac{2J}{J+\mu} e^{-\mathcal{S}(x)}\Bigl [ 1+ i \frac{4(l-1)^2-1}{8x}\Bigr ] 
\end{align}
Taking into account the asymptotic expansion \eqref{eq:Jn:a} for the Bessel function, we 
choose $\tilde{c}_0= -c_0 e^{-i\pi/4-i\pi(l+1)/2}/\sqrt{2\pi}$ and, then, can approximate the solutions for $u_\downarrow$ and $v_\downarrow$ as in Eqs.~\eqref{eq:down:WKB:lnot0}.

Next, from Eqs. \eqref{eq:DD:1}-\eqref{eq:DD:2}, we obtain
\begin{align}
A_{\uparrow} & \simeq  (\sgn\a) \tilde{c}_0 e^{{-}\mathcal{S}(x)}\Bigl [ i|\bar{\alpha}| \Bigl(1{+} i \frac{4l^2{-}1}{8x}\Bigr ) \notag \\
& {\,}\hspace{2.7cm} {-} \bar{\Delta}\Bigl ( 1{+} i \frac{4(l{-}1)^2{-}1}{8x}\Bigr ) \Bigr ]  ,  \\
B_{\uparrow} & \simeq  {-}\tilde{c}_0 e^{{-}\mathcal{S}(x)}\Bigl [ i|\bar{\alpha}| \Bigl(1{+} i \frac{4l^2{-}1}{8x}\Bigr ) {-} \bar{\Delta}\Bigl ( 1{+} i \frac{4(l{+}1)^2{-}1}{8x}\Bigr ) \Bigr ] 
\label{eq:DD:14}.
\end{align}
Again, using the asymptotic expansion \eqref{eq:Jn:a}, we construct the solutions for $ u_\uparrow$ and $v_\uparrow$ in terms of the Bessel functions, see Eqs.~\eqref{eq:up:WKB:lnot0}. \color{black}
We note that strictly speaking the asymptotic solutions \eqref{eq:DD:11}-\eqref{eq:DD:14} works at $x\gg\max(1,|l|)$.
\color{black}

Using the normalization condition,
\begin{equation}
  \int d^2 \bm{r} \Bigl [u_{\uparrow,l}^{v2}+u_{\downarrow,l}^{v2}+v_{\uparrow,l}^{v2}+v_{\downarrow,l}^{v2}\Bigr ] = 1,  
\end{equation}
we find 
\begin{gather}
 c_{v,l}^{-2} \simeq 8\pi \left (\frac{R_J^2 J}{\alpha}\right )^2 \int_0^\infty dx x e^{-2\mathcal{S}(r)}\Bigl [\mathcal{J}^2_{l-1}(x)+\mathcal{J}^2_{l+1}(x)\Bigr ]
 \notag \\
 \simeq 16\pi \left (\frac{R_J^2 J}{\alpha}\right )^2 \int_0^\infty dx x \mathcal{J}^2_{l}(x) e^{-2\mathcal{S}(r)} .
 \label{eq:app:cvl:norm}
\end{gather}
In order to estimate the eigen energies $E_l$ we compute the average value of the Hamiltonian $H^{(l)}$ with the eigen function $\Phi_l$, given by Eqs. \eqref{eq:down:WKB:lnot0}-\eqref{eq:up:WKB:lnot0}. We find 
\begin{gather}
 \langle \Phi_l | H^{(l)}|\Phi_l\rangle  \simeq - 16\pi lc_{v,l}^2 \frac{R^3_J J}{|\alpha|}
   \int\limits_0^\infty dx \Delta(r)
   \Bigl [ \mathcal{J}_l^2(x) \notag \\
   -\frac{1}{2x} \partial_x \mathcal{J}_l^2(x)\Bigr ] e^{-2\mathcal{S}(r)}
\end{gather}
We omit the term in the second line of the above equations since it provides a correction of the order of $1$ in comparison with logarithm (see Eq.~\eqref{eq:El:main}). Then we obtain
\begin{gather}
E_l^v \simeq - \frac{|\alpha| l}{R_J J}    \frac{\int_0^\infty dx \Delta e^{-2\mathcal{S}(r)}\mathcal{J}_l^2(x)}{\int_0^\infty dx x e^{-2\mathcal{S}(r)}\mathcal{J}_l^2(x) }
\end{gather}
Evaluating the integral, we obtain Eq. \eqref{eq:El:main} in the main text. We note that in order to estimate the energies $E_l$ once can use other quantities. For example, one can compute the average of $d H^{(l)}/d\Delta_\infty$ which is equal to $dE_l/d\Delta_\infty$ in virtue of the Hellmann-Feynman theorem. As one can check, it gives the same dependence of the energy $E_l$ on the parameters but with numerical coefficient 2 times smaller than in Eq. \eqref{eq:El:main}.

\subsection{The case of the presence of a skyrmion}

Now we construct solution for the wave functions for the state with non-zero $l$ localized on a superconducting vortex with a coaxial skyrmion. We employ the unitary transformation \eqref{eq:eq:M:R} to the Hamiltonian $H^{(l)}$, cf. Eq. \eqref{eq:Hl}. 
Then we obtain
\begin{widetext}
    \begin{align}
        \underline{H}^{(l)}&=-\frac{\tau_z}{2mR_J^2}\left (\partial_x^2+\frac{1}{x}\partial_x\right )+\frac{\tau_z}{2mR_J^2x^2}\Big(l+\frac{1}{2}\big(\tau_z-\sigma_z\cos\vartheta+\sigma_x\sin\vartheta\big)\Big)^2-\mu\tau_z+\D \tau_x -J\sigma_z+i\tau_z\sigma_y\frac{\partial_x^2\vartheta}{4mR_J^2}+ \notag \\
        &+\frac{i\tau_z\sigma_y}{R_J}\Big(\a+\frac{\pd_x\vartheta}{2mR_J}\Big)\Big(\pd_x+\frac{1}{2x}\Big)+\tau_z\frac{(\pd_x\vartheta)^2}{8mR_J^2}+\tau_z\frac{\a\pd_x\vartheta}{2R_J}+\frac{\a}{2R_Jx}\Big(1+2l\tau_z\Big)\Big(\sigma_x\cos{\vartheta}+\sigma_z\sin{\vartheta}\Big) .
    \end{align}
By neglecting the spatial derivatives of $\vartheta$, we find that the 4-component wave function is determined by the following system of equations: 
		\begin{gather}
			\partial_x^2 \underline{u}_{\uparrow}+\frac{1}{x}\partial_x \underline{u}_\uparrow-\frac{1}{x^2}\Big((l+\frac{1}{2}(1-\cos\vartheta))^2+\frac{1}{4}\sin^2\vartheta\Big)\underline{u}_\uparrow+\underline{u}_\uparrow-\frac{\bar{\a}(1+2l)}{2x}\sin\vartheta \underline{u}_\uparrow+\bar{E}_l \underline{u}_\uparrow
            \notag
            \\
            -\frac{\sin\vartheta(l+1/2)}{x^2}\underline{u}_\downarrow-\bar{\D}\underline{v}_\downarrow-\bar{\a}\Big(\partial_x\underline{u}_\downarrow+\frac{\underline{u}_\downarrow}{2x}\Big)-\frac{\bar{\a}(1+2l)}{2x}\cos\vartheta \underline{u}_\downarrow=0 ,
            \label{eq:D1Sk}
            \end{gather}
            \begin{gather}
			\partial_x^2 \underline{u}_{\downarrow}+\frac{1}{x}\partial_x\underline{u}_\downarrow-\frac{1}{x^2}\Big((l+\frac{1}{2}(1+\cos\vartheta))^2+\frac{1}{4}\sin^2\vartheta\Big)\underline{u}_\downarrow-\frac{J-\mu}{J+\mu} \underline{u}_\downarrow+\frac{\bar{\a}(1+2l)}{2x}\sin\vartheta \underline{u}_\downarrow+\bar{E}_l \underline{u}_\downarrow\notag \\
            -\frac{\sin\vartheta(l+1/2)}{x^2} \underline{u}_\uparrow+\bar{\D} \underline{v}_\uparrow+\bar{\a}\Big(\partial_x \underline{u}_\uparrow+\frac{\underline{u}_\uparrow}{2x}\Big)-\frac{\bar{\a}(1+2l)}{2x}\cos\vartheta \underline{u}_\uparrow=0 , \label{eq:D2Sk}\\
            \partial_x^2 \underline{v}_{\uparrow}+\frac{1}{x}\partial_x \underline{v}_\uparrow-\frac{1}{x^2}\Big((l-\frac{1}{2}(1-\cos\vartheta))^2+\frac{1}{4}\sin^2\vartheta\Big)\underline{v}_\uparrow+\underline{v}_\uparrow-\frac{\bar{\a}(1-2l)}{2x}\sin\vartheta \underline{v}_\uparrow-\bar{E}_l\underline{v}_\uparrow\notag \\
            +\frac{\sin\vartheta(l-1/2)}{x^2}\underline{v}_\downarrow-\bar{\D}\underline{u}_\downarrow-\bar{\a}\Big(\partial_x \underline{v}_\downarrow+\frac{\underline{v}_\downarrow}{2x}\Big)-\frac{\bar{\a}(1-2l)}{2x}\cos\vartheta \underline{v}_\downarrow= 0 , 
            \label{eq:D3Sk}
			\\
			\partial_x^2 \underline{v}_{\downarrow}+\frac{1}{x}\partial_x \underline{v}_\downarrow-\frac{1}{x^2}\Big((l-\frac{1}{2}(1+\cos\vartheta))^2+\frac{1}{4}\sin^2\vartheta\Big)\underline{v}_\downarrow-\frac{J-\mu}{J+\mu} \underline{v}_\downarrow +\frac{\bar{\a}(1-2l)}{2x}\sin\vartheta \underline{v}_\downarrow-\bar{E}_l\underline{v}_\downarrow\notag\\
            +\frac{\sin\vartheta(l-1/2)}{x^2}\underline{v}_\uparrow+\bar{\D}\underline{u}_\uparrow+\bar{\a}\Big(\partial_x \underline{v}_\uparrow+\frac{\underline{v}_\uparrow}{2x}\Big)-\frac{\bar{\a}(1-2l)}{2x}\cos\vartheta \underline{v}_\uparrow= 0 .
             \label{eq:D4Sk}
		\end{gather}
Now we substitute $\underline{u}_{\uparrow,\downarrow}$ and $\underline{v}_{\uparrow,\downarrow}$ in the form of Eq. \eqref{eq:u:v:asymp}. Then we obtain the following exact equations for $\underline{A}_{\uparrow,\downarrow}$ and $\underline{B}_{\uparrow,\downarrow}$:
    \begin{gather}
        	\partial_x^2 \underline{A}_{\uparrow}+2i \partial_x \underline{A}_{\uparrow}+\frac{\underline{A}_{\uparrow}}{4x^2}-\frac{1}{x^2}\Big((l+\frac{1}{2}(1-\cos\vartheta))^2+\frac{1}{4}\sin^2\vartheta\Big)\underline{A}_\uparrow-\frac{\bar{\a}(1+2l)}{2x}\sin\vartheta \underline{A}_\uparrow+\bar{E}_l\underline{A}_\uparrow \notag \\
            -\frac{\sin\vartheta(l+1/2)}{x^2}\underline{A}_\downarrow-\bar{\D}\underline{B}_\downarrow-\bar{\a}\Big(\partial_x \underline{A}_\downarrow+i\underline{A}_{\downarrow}\Big)-\frac{\bar{\a}(1+2l)}{2x}\cos\vartheta \underline{A}_\downarrow=0 ,
            \label{eq:D1Sk}
            \\
            \partial_x^2 \underline{A}_{\downarrow}+2i\partial_x \underline{A}_{\downarrow}+\frac{\underline{A}_{\downarrow}}{4x^2}-\frac{1}{x^2}\Big((l+\frac{1}{2}(1+\cos\vartheta))^2+\frac{1}{4}\sin^2\vartheta\Big)\underline{A}_\downarrow-\frac{2J}{J+\mu} \underline{A}_\downarrow+\frac{\bar{\a}(1+2l)}{2x}\sin\vartheta \underline{A}_\downarrow+\bar{E}_l\underline{A}_\downarrow \notag \\
            -\frac{\sin\vartheta(l+1/2)}{x^2}\underline{A}_\uparrow+\bar{\D}\underline{B}_\uparrow+\bar{\a}\Big(\partial_x \underline{A}_\uparrow+i\underline{A}_{\uparrow}\Big)-\frac{\bar{\a}(1+2l)}{2x}\cos\vartheta \underline{A}_\uparrow=0 , \label{eq:D2Sk}
            \end{gather}
            \begin{gather}
            \partial_x^2 \underline{B}_{\uparrow}+2i\partial_x \underline{B}_{\uparrow}+\frac{\underline{B}_{\uparrow}}{4x^2}-\frac{1}{x^2}\Big((l-\frac{1}{2}(1-\cos\vartheta))^2+\frac{1}{4}\sin^2\vartheta\Big)\underline{B}_\uparrow-\frac{\bar{\a}(1-2l)}{2x}\sin\vartheta \underline{B}_\uparrow-\bar{E}_l\underline{B}_\uparrow \notag \\
            +\frac{\sin\vartheta(l-1/2)}{x^2}\underline{B}_\downarrow-\bar{\D}\underline{A}_\downarrow-\bar{\a}\Big(\partial_x \underline{B}_\downarrow+i\underline{B}_{\downarrow}\Big)-\frac{\bar{\a}(1-2l)}{2x}\cos\vartheta \underline{B}_\downarrow=0 ,
            \label{eq:D3Sk}
			\\
			\partial_x^2 \underline{B}_{\downarrow}+2i \partial_x \underline{B}_{\downarrow}+\frac{\underline{B}_{\downarrow}}{4x^2}-\frac{1}{x^2}\Big((l-\frac{1}{2}(1+\cos\vartheta))^2+\frac{1}{4}\sin^2\vartheta\Big)\underline{B}_\downarrow-\frac{2J}{J+\mu} \underline{B}_\downarrow+\frac{\bar{\a}(1-2l)}{2x}\sin\vartheta \underline{B}_\downarrow-\bar{E}_l\underline{B}_\downarrow\notag \\
            +\frac{\sin\vartheta(l-1/2)}{x^2}\underline{B}_\uparrow+\bar{\D}\underline{A}_\uparrow+\bar{\a}\Big(\partial_x \underline{B}_\uparrow+i\underline{B}_{\uparrow}\Big)-\frac{\bar{\a}(1-2l)}{2x}\cos\vartheta \underline{B}_\uparrow=0 .
             \label{eq:D4SkA} 
    \end{gather}
\end{widetext}
As before, we neglect $\bar{E}_l$ and assume $\underline{A}_{\uparrow,\downarrow}$ and $\underline{B}_{\uparrow,\downarrow}$ to be slow functions. Then we obtain the following relations:
\begin{gather}
 \underline{A}_\downarrow=\frac{J+\mu}{2J} \Bigl [ 
            \bar{\D}\underline{B}_\uparrow+\Bigl (i\bar{\a}-\frac{\bar{\a}(1+2l)}{2x}\cos\vartheta \Bigr )\underline{A}_\uparrow\Bigr ] , \label{eq:Add:1} \\
  \underline{B}_\downarrow=\frac{J+\mu}{2J} \Bigl [ \Bigl (i\bar{\a}-\frac{\bar{\a}(1-2l)}{2x}\cos\vartheta \Bigr )\underline{B}_\uparrow +\bar{\D}\underline{A}_\uparrow\Bigr ] ,        \label{eq:Bdd:1}
\end{gather}
and the following first order differential equations: 
\begin{align}
2i \partial_x \underline{A}_{\uparrow} = & (2l{+}1)(2l{+}1{-}2\cos\vartheta)\frac{\underline{A}_{\uparrow}}{4x^2}+\frac{\bar{\a}(2l{+}1)}{2x}\sin\vartheta \underline{A}_\uparrow\notag \\
& +\bar{\D}\underline{B}_\downarrow+i \bar{\a}\underline{A}_{\downarrow}  , 
\label{eq:Auu:1} \\
2i\partial_x \underline{B}_{\uparrow}
= & (2l{-}1)(2l{-}1{+}2\cos\vartheta)
\frac{\underline{B}_{\uparrow}}{4x^2}+\frac{\bar{\a}(1{-}2l)}{2x}\sin\vartheta \underline{B}_\uparrow\notag \\ & +\bar{\D}\underline{A}_\downarrow+i \bar{\a}\underline{B}_{\downarrow} .
            \label{eq:B:uu:1}
\end{align}
Solving the system of equations \eqref{eq:Add:1}-\eqref{eq:B:uu:1}, we find
\begin{align}
 \underline{A}_{\uparrow} & {=}\frac{2J b_l}{J+\mu} e^{-\mathcal{S}(r)}
 \Bigl [ 1{+}i \frac{(2l{+}1)(2l{+}1{-}2\cos\vartheta)}{8 x} \Bigr ], \label{eq:A:uu:2}\\
 \underline{B}_{\uparrow} & {=} {-}(\sgn\a) \frac{2J b_l}{J+\mu} e^{-\mathcal{S}(r)}
 \Bigl [ 1{+}i \frac{(2l{-}1)(2l{-}1{+}2\cos\vartheta)}{8 x} \Bigr ], \label{eq:B:uu:2}
\end{align}
and
\begin{align}
 \underline{A}_{\downarrow} & {=}{-}(\sgn\a)b_l e^{-\mathcal{S}(r)}
 \Biggl [ \Bigl( 1{+}i \frac{(2l{-}1)(2l{-}1{+}2\cos\vartheta)}{8 x}\Bigr)\bar{\Delta} \notag \\
 &{-} i \Bigl( 1{+}i \frac{(2l{+}1)(2l{+}1{+}2\cos\vartheta)}{8 x}\Bigr)|\bar{\alpha}|\Biggr ], \label{eq:A:dd:2}
 \end{align}
 \begin{align}
 \underline{B}_{\downarrow} & {=} b_l e^{-\mathcal{S}(r)}\Biggl [
 \Bigl( 1{+}i \frac{(2l{+}1)(2l{+}1{-}2\cos\vartheta)}{8 x}\Bigr)\bar{\Delta} \notag \\
 &{-} i\Bigl( 1{+}i \frac{(2l{-}1)(2l{-}1{-}2\cos\vartheta)}{8 x}\Bigr)|\bar{\alpha}|\Biggr ], \label{eq:B:dd:2}
\end{align}
Taking into account the asymptotic expressions for the Bessel functions, Eqs. \eqref{eq:Jn:a} and \eqref{eq:Yn:a}, we choose the constant $b_l= -(\tilde{c}_l/(\bar{\alpha}\sqrt{2\pi}))\exp(-i\pi/4-i\pi l/2)$ and, then, construct the solutions as
\begin{equation}
 \underline{u}_\uparrow  \simeq  {-}\frac{2 \tilde{c}_l R_J J} {\alpha} e^{{-}\mathcal{S}(r)}\Bigl [
\mathcal{J}_lx) \cos^2\frac{\vartheta}{2}  {-} \mathcal{Y}_{l+1}(x) \sin^2\frac{\vartheta}{2} \Bigr ] \label{eq:u:uu:2}
\end{equation}
\begin{equation}
\underline{v}_\uparrow  \simeq  \frac{2 \tilde{c}_l R_J J} {|\alpha|} e^{{-}\mathcal{S}(r)}\Bigl [
\mathcal{J}_l(x) \cos^2\frac{\vartheta}{2}  {+} \mathcal{Y}_{l-1}(x) \sin^2\frac{\vartheta}{2} \Bigr ] \label{eq:v:uu:2}
\end{equation}
and
\begin{align}
\underline{u}_\downarrow = &  \tilde{c}_l e^{{-}\mathcal{S}(r)}   
\Biggl [ \mathcal{J}_{l+1}(x) \cos^2\frac{\vartheta}{2} + \mathcal{Y}_{l}(x) \sin^2\frac{\vartheta}{2} \notag \\
& + \frac{\Delta R_J}{|\alpha|} \Bigl( \mathcal{J}_{l}(x) \cos^2\frac{\vartheta}{2} + \mathcal{Y}_{l-1}(x) \sin^2\frac{\vartheta}{2}\Bigr )
\Biggr ] , \label{eq:u:dd:2} 
\end{align}
\begin{align}
\underline{v}_\downarrow = & -(\sgn\a) \tilde{c}_l e^{{-}\mathcal{S}(r)}   
\Biggl [ -\mathcal{J}_{l-1}(x) \cos^2\frac{\vartheta}{2}+ \mathcal{Y}_{l}(x) \sin^2\frac{\vartheta}{2}  \notag \\
& +  \frac{\Delta R_J}{|\alpha|} \Bigl( \mathcal{J}_{l}(x)  \cos^2\frac{\vartheta}{2} - \mathcal{Y}_{l+1}(x)  \sin^2\frac{\vartheta}{2}\Bigr )
\Biggr ] .\label{eq:v:dd:2}
\end{align}
We note that for $x\gtrsim 10$, we can use the following approximate relations: $\mathcal{Y}_{l\pm1}(x)\simeq \mp J_l(x)$ and $\mathcal{Y}_l(x)\simeq \pm J_{l\pm 1}(x)$. Then Eqs. \eqref{eq:u:uu:2} - \eqref{eq:v:dd:2} simplifies to Eqs.~\eqref{eq:v:dd:3}.

Similar to the case  without skyrmion, in order to estimate the energy $E_l$ of the states with nonzero $l$, we compute the average of the rotated Hamiltonian $\underline{H}^{(l)}$ over the wave functions \eqref{eq:v:dd:3}. We note that we do not use the expressions \eqref{eq:u:uu:2}-\eqref{eq:v:dd:2} with the Bessel functions of the second kind. After tedious but straightforward calculations, we find
\begin{gather}
\langle \underline{\Phi}_{l}|\underline{H}^{(l)}|\underline{\Phi}_{l}\rangle \simeq 
\frac{8\pi l c_{v,l}^2}{m} \left (\frac{R_J J}{\alpha}\right )^2 \int_0^\infty dx e^{-2\mathcal{S}(r)}\mathcal{J}_l^2(x)\notag \\ \times \Biggl [ \frac{1-\cos\vartheta}{x} +\bar{\alpha} \sin \vartheta-\frac{\Delta(\sgn\a)\sin \vartheta}{J x}- \frac{|\bar{\alpha}|\Delta \cos\vartheta}{J} \Bigr ] ,
\end{gather}
where normalization constant $c_{v,l}$ is given by Eq.~\eqref{eq:app:cvl:norm}. Using the above result, we obtain Eq. \eqref{eq:El:WKB:Sk} in the main text.

\bibliography{bib-skyrmion,skyrmion2}

\end{document}